\documentclass[onecolumn,amsmath,amssymb,11pt]{revtex4}
\usepackage{psfrag}
\usepackage{epsfig}
\usepackage{graphicx,graphics}
 \normalsize

\def\lsim{\raise0.3ex\hbox{$\;<$\kern-0.75em\raise-1.1ex\hbox{$\sim\;$}}}
\def\gsim{\raise0.3ex\hbox{$\;>$\kern-0.75em\raise-1.1ex\hbox{$\sim\;$}}}

\def\2tvec#1#2{ \left( \begin{array}{c}
#1  \\
#2  \\
\end{array} \right)}%
\def\mat2#1#2#3#4{ \left( \begin{array}{cc}
#1 & #2 \\
#3 & #4 \\
\end{array} \right) }%
\def\Mat3#1#2#3#4#5#6#7#8#9{ \left( \begin{array}{ccc}tri-bimaximal
#1 & #2 & #3 \\
#4 & #5 & #6 \\
#7 & #8 & #9 \\
\end{array} \right) } %
\def\Mat3#1#2#3#4#5#6#7#8#9{ \left(
\begin{array}{ccc}
#1 & #2 & #3 \\
#4 & #5 & #6 \\
#7 & #8 & #9 \\
\end{array} \right) }

\def\3tvec#1#2#3{ \left( \begin{array}{c}
#1  \\
#2  \\
#3  \\
\end{array} \right)}

\def\4tvec#1#2#3#4{ \left( \begin{array}{c}
#1  \\
#2  \\
#3  \\
#4  \\
\end{array} \right)}

\def\hbar{\hspace{1mm}\bar{}\hspace{-1mm}h}

\def\bea{\begin{eqnarray}}
\def\eea{\end{eqnarray}} \newcommand{\be}{\begin{eqnarray}}
\newcommand{\ee}{\end{eqnarray}}
\usepackage{latexsym}
\usepackage{graphicx} \usepackage{verbatim}
\begin{document}
\title{Lepton masses and mixing in non-holomorphic modular $A_4$ with universal couplings}
\author{
{\textbf{Mohammed Abbas}}\thanks{email: \tt maabbas@ju.edu.sa}
 \\
{\normalsize\em
Physics Department, College of Science, Jouf University, Sakaka, P.O.Box 2014,} \\
{\normalsize\em  Saudi Arabia.
\vspace*{0.15cm}}
}
\maketitle

We propose a non-holomorphic modular $A_4$ model under the assumption of universal couplings. In this framework, a charged lepton mass hierarchy is not created through parameter fine tuning or hierarchical Yukawa couplings, but instead is determined by the modulus $\tau$, with certain modular weight assignments of right handed charged leptons. The experimental charged lepton masses are reproduced with high precision for values of $\tau$ located near modular fixed points. In neutrino sector the couplings are imposed to be equal in magnitude with different relative phases. By fixing the modulus $\tau$ from charged lepton sector, we perform a comprehensive scan over the phase parameters and modular weight assignments of the right handed neutrinos. We find that viable solutions arise only for normal neutrino mass ordering and a unique right handed neutrino modular weight, $k_N=-1$. 
The model yields strong correlations among mixing angles, the effective neutrinoless double beta decay parameter $m_{ee}$, and the total neutrino mass $\sum m_i$. These results underscore the predictive quality of non-holomorphic modular symmetry with minimal parameter inputs and offer implications for neutrino experiments and cosmological observations that can be tested.

\linespread{1.2}


\maketitle

\vspace*{10mm}


\fontsize{10}{11}\selectfont

\section{Introduction}

The origin of fermion masses and flavor mixing is still a problem in particle physics.
In the lepton sector the mixing pattern is very different from quarks.
This suggests that its structure might be due to an underlying symmetry, not random choices of parameters.

In most of conventional flavor models, the hierarchy of charged lepton masses is typically attributed to fine tuned hierarchical Yukawa couplingsa. While this approach is phenomenologically successful, it does not address the origin of these hierarchies. An alternative viewpoint is to consider that the Yukawa couplings in charged lepton sector are universal, with the observed hierarchy emerging from the modular structure, specifically from the assignment of modular weights and the value of the modulus $\tau$. Consequently, the hierarchy has a geometric origin rather than being imposed through parameter tuning.

In modular flavor models, Yukawa couplings are promoted to modular forms depending on a complex modulus $\tau$, and fermion properties are determined by modular weights and discrete symmetry representations. This construction significantly reduces the number of free parameters and leads to correlations among physical observables \cite{modulargroups, Feruglio:2017spp}.

An important extension of this idea is the non-holomorphic modular framework, in which supersymmetry is not required and Yukawa couplings are described by polyharmonic Maa\ss\ forms \cite{Qu:2024rns}. This generalization allows modular forms to carry positive, zero and negative weights, thereby enlarging the space of viable models while maintaining predictivity. Several non-holomorphic modular symmetries have been explored, including non-holomorphic $S_3$ \cite{Okada:2025jjo}, non-holomorphic $A_4$ \cite{Loualidi:2025tgw, Abbas:2025nlv, Nomura:2025raf, Zhang:2025dsa, Priya:2025wdm, Kumar:2025nut, Nanda:2025lem, Jangid:2025thp, Gao:2025jlw, Tapender:2026ets, Majhi:2026jdk, Nomura:2024atp, Nomura:2024vzw}, non-holomorphic $S_4$ \cite{Ding:2024inn}, and non-holomorphic $A_5$ \cite{Li:2024svh, Zhang:2026kyy, Li:2025kcr}.

In this work, we explore a non-holomorphic modular $A_4$ model under the assumption that all couplings in charged lepton sector have the same magnitude. A comprehensive scan over the discrete values of the modular weights of right handed charged leptons and the modulus $\tau$ in the vicinity of some fixed points is performed to achieve the observed charged lepton masses. This implies that the mass hierarchy is entirely determined by the modulus $\tau$ and the assignment of modular weights. Therefore, the model is free from the problem of fine tuning of couplings and the coupling hierarchies. By fitting the charged lepton masses, the modulus $\tau$ is effectively fixed, the freedom of the model is reduced.

The neutrino sector is then constructed via the type-I seesaw mechanism. We assume the couplings in neutrino sector to have the same amplitude with different phases. We perform a systematic scan over phase parameters and right handed neutrino modular weight. A numerical analysis leads to nontrivial predictions for neutrino observables and reveals that the model is highly constrained. Viable solutions are found only for normal ordering (NO) and a unique modular weight $k_N=-1$. The Model provides strong correlations among mixing angles, the effective neutrinoless double beta decay parameter $m_{ee}$, and the sum of neutrino masses $\sum m_i$.

A notable feature of the framework is the emergence of charged lepton permutation selection. Different orderings of charged lepton eigenstates lead to distinct physical predictions, and only a subset of these is consistent with experimental data. This indicates that the relative alignment between the charged lepton and neutrino sectors is dynamically determined.

The paper is organized as follows. In Sec.~\ref{sec:Non-holomorphic_symmetry}, we review the non-holomorphic modular symmetry. In Sec.~\ref{sec:model}, we describe the model setup. Numerical results are presented in Sec.~\ref{sec:results}, followed by a discussion in Sec.~\ref{sec:discussion}. We conclude in Sec.~\ref{sec:conclusion}.

\section{Non-holomorphic modular flavor symmetry} \label{sec:Non-holomorphic_symmetry}
In this section, we provide a brief overview of the non-holomorphic modular flavor symmetry. The homogeneous modular group $\Gamma \equiv SL(2,\mathbb{Z})$ consists of $2\times2$ matrices with integer entries and unit determinant. The projective modular group is defined as
\bea
PSL(2,\mathbb{Z}) = SL(2,\mathbb{Z})/\{I,-I\},
\eea
where
\bea
SL(2,\mathbb{Z}) = \left\{
\begin{pmatrix}
a & b \\
c & d
\end{pmatrix}
\,\bigg|\, a,b,c,d \in \mathbb{Z}, \; ad - bc = 1
\right\}.
\eea

The group $\Gamma$ is generated by two elements $S$ and $T$, which can be represented as
\begin{equation}
S =
\begin{pmatrix}
0 & 1 \\
-1 & 0
\end{pmatrix}, \qquad
T =
\begin{pmatrix}
1 & 1 \\
0 & 1
\end{pmatrix}.
\end{equation}
These generators satisfy the defining relations
\begin{equation}
S^2 = \mathbf{1}, \qquad (ST)^3 = \mathbf{1}.
\label{st_condition_re}
\end{equation}

The action of the modular group on the complex upper half-plane $\mathcal{H}$ is given by
\bea
\gamma: \tau \rightarrow \gamma(\tau) = \frac{a\tau + b}{c\tau + d},
\label{gamma_re}
\eea
where $\tau \in \mathcal{H}$ and $a,b,c,d \in \mathbb{Z}$ with $ad-bc=1$. In particular, the generators act as
\bea
S: \tau \rightarrow -\frac{1}{\tau}, \qquad T: \tau \rightarrow \tau + 1.
\eea

The infinite modular subgroups $\Gamma(N)$ are defined by
\bea
\Gamma(N) = \left\{
\begin{pmatrix}
a & b \\
c & d
\end{pmatrix} \in SL(2,\mathbb{Z}) \;\bigg|\;
\begin{pmatrix}
a & b \\
c & d
\end{pmatrix}
\equiv
\begin{pmatrix}
1 & 0 \\
0 & 1
\end{pmatrix}
\; (\mathrm{mod}\, N)
\right\}.
\label{GammaN_re}
\eea

For $N=1$, one trivially recovers $\Gamma(1) = SL(2,\mathbb{Z})$. For $N=1,2$, one defines $\bar{\Gamma}(N) = \Gamma(N)/\{I,-I\}$, while for $N>2$ one has $\bar{\Gamma}(N) = \Gamma(N)$ since $-I \notin \Gamma(N)$. The quotient group
\begin{equation}
\Gamma_N = \bar{\Gamma}/\bar{\Gamma}(N)
\end{equation}
is finite and is referred to as the finite modular group. These groups are isomorphic to well-known permutation groups for specific values of $N$, such as $\Gamma_2 \cong S_3$, $\Gamma_3 \cong A_4$, $\Gamma_4 \cong S_4$, and $\Gamma_5 \cong A_5$.

Modular flavor symmetry was originally formulated in supersymmetric frameworks, where holomorphicity constrains Yukawa couplings to be modular forms. More recently, a non-holomorphic extension has been proposed~\cite{Qu:2024rns}, in which supersymmetry is not required. In this approach, Yukawa couplings are described by polyharmonic Maa{\ss} forms of level $N$.

These Maa{\ss} forms, denoted by $Y(\tau)$, carry a modular weight $k$ and transform under the modular group as
\begin{equation}
Y(\tau) \rightarrow Y(\gamma\tau) = (c\tau + d)^k Y(\tau), \qquad
\gamma \in \Gamma(N).
\end{equation}
They satisfy the differential equation
\begin{eqnarray}
\nonumber
&&\left(-4y^2 \frac{\partial}{\partial \tau} \frac{\partial}{\partial \bar{\tau}} + 2iky \frac{\partial}{\partial \bar{\tau}}\right) Y(\tau) = 0, \\
&&Y(\tau) = \mathcal{O}(y^{\alpha}) \quad \text{as } y \rightarrow \infty,
\end{eqnarray}
where $k$ can take positive, zero, or negative integer values.

Matter fields transform under the modular symmetry as
\begin{eqnarray}
\nonumber
\psi(x) &\rightarrow& (c\tau + d)^{-k_\psi} \rho_\psi(\gamma)\, \psi(x), \\
\nonumber
\psi^c(x) &\rightarrow& (c\tau + d)^{-k_{\psi^c}} \rho_{\psi^c}(\gamma)\, \psi^c(x), \\
H(x) &\rightarrow& (c\tau + d)^{-k_H} \rho_H(\gamma)\, H(x),
\end{eqnarray}
where $\rho(\gamma)$ denotes a unitary representation of the finite modular group.

The Yukawa Lagrangian can then be written as
\begin{equation}
\mathcal{L}_Y = -Y^{(k_Y)}(\tau)\, \psi^c \psi H + \mathrm{h.c.},
\end{equation}
with $H \to H^*$ for down-type fermions. The modular transformation of the field coupling $Y^{(k_Y)}(\tau)$ is expressed in the following manner:

\begin{equation}
Y^{(k_Y)}(\tau)\mapsto Y^{(k_Y)}(\gamma\tau)=(c\tau+d)^{k_Y}\rho_{Y}(\gamma)Y^{(k_Y)}(\tau)\,,
\end{equation}
Modular invariance requires that the modular weights and representations satisfy
\begin{equation}
k_Y = k_{\psi^c} + k_\psi + k_H, \qquad
\rho_Y \otimes \rho_{\psi^c} \otimes \rho_\psi \otimes \rho_H \ni \mathbf{1}.
\label{modular_condition_re}
\end{equation}

\subsection{Polyharmonic Maa{\ss} forms of level 3}

The group $A_4$ is generated by $S$ and $T$, satisfying
\[
S^2 = T^3 = (ST)^3 = \mathbf{1}.
\]
In the triplet representation, they can be written as
\bea
S = \frac{1}{3}
\begin{pmatrix}
-1 & 2 & 2 \\
2 & -1 & 2 \\
2 & 2 & -1
\end{pmatrix}, \qquad
T =
\begin{pmatrix}
1 & 0 & 0 \\
0 & \omega & 0 \\
0 & 0 & \omega^2
\end{pmatrix}.
\eea

For level $N=3$, the polyharmonic Maa{\ss} forms can be classified according to their modular weight $k$ in the range $-4 \le k \le 6$ as follows:
\begin{itemize}
\item For $k>2$, they coincide with the holomorphic modular forms of $\Gamma(3)$.
\item For $k=2$, the independent forms are $Y_3^{(2)}$ and $Y_1^{(2)}$.
\item For $k=0$, the independent forms are $Y_3^{(0)}$ and $Y_1^{(0)}$.
\item For $k=-2$, the independent forms are $Y_3^{(-2)}$ and $Y_1^{(-2)}$.
\item For $k=-4$, the independent forms are $Y_3^{(-4)}$ and $Y_1^{(-4)}$.
\end{itemize}

For $k \le 2$, the space of Maa{\ss} forms has dimension four, consisting of one triplet and one singlet. Further details on their explicit construction can be found in Ref.~\cite{Qu:2024rns}.

\subsection{Modular residual symmetry}

The modular group admits a set of special points in the fundamental domain, commonly referred to as fixed points, which are invariant under specific modular transformations. These points are given by
$ e^{\frac{2\pi i}{3}} = -\frac{1}{2} + i\frac{\sqrt{3}}{2}, ~
 i, ~
 i\infty.
$

Each of these points remains unchanged under a nontrivial element of the modular group. Any other fixed point can be related to one of the above through modular transformations. For example, applying the transformation $T$ to $-\frac{1}{2} + i\frac{\sqrt{3}}{2}$ yields the equivalent point
$
\frac{1}{2} + i\frac{\sqrt{3}}{2}.
$
Similarly, acting with $ST$ on $i$ leads to $-\frac{1}{2} + \frac{i}{2}$, while a further transformation $TST$ generates $ \frac{1}{2} + \frac{i}{2}$. 

A key property of these points is that the full modular symmetry is not preserved but instead reduced to a residual subgroup. This residual symmetry plays an important role in shaping the structure of fermion mass matrices. In particular, the point $e^{\frac{2\pi i}{3}}$ is invariant under the action of $ST$, leading to the breaking of $A_4$ down to the subgroup $Z_3 = \{I, ST, (ST)^2\}$. Consequently, the fermion mass matrices exhibit invariance under the $ST$ transformation.

On the other hand, the point $i$ remains invariant under the generator $S$, resulting in the reduction of $A_4$ to the subgroup $Z_2 = \{I, S\}$. In this case, the corresponding mass matrices respect the $S$ symmetry. The point $-\frac{1}{2} + \frac{i}{2}$ is left invariant by the transformation $ST^2ST$, which leads to a different $Z_2$ residual subgroup. Finally, the cusp $ i\infty$ is invariant under $T$, implying a residual $Z_3 = \{I, T, T^2\}$ symmetry.

In this framework, we analyze the structure of charged lepton masses in the vicinity of modular fixed points. These special points in the modulus space correspond to enhanced residual symmetries, which can significantly constrain the form of the mass matrices. By exploring regions close to these fixed points, we investigate how small deviations from exact modular symmetry can generate the observed charged lepton mass hierarchy within the framework of universal couplings.

\section{$A_4$ non-holomorphic modular invariance model}\label{sec:model}
In modular flavor frameworks, the effective Yukawa couplings can be considered as combinations of modular forms $Y(\tau)$ and modular invariant coefficients. In the present model, we adopt the simplifying assumption that these invariant couplings in the charged lepton sector are universal having exactly the same values. Under this assumption, the charged lepton mass hierarchy is generated by assigning different modular weights to the modular forms associated with each fermion generation. In this case, the hierarchy does not originate from ad hoc Yukawa parameters, but instead emerges from the underlying modular structure. In this sense, the differences between generations are governed by the modular geometry rather than by arbitrary choices of coupling constants.

Our analysis focuses on regions close to modular fixed points, where the structure of the theory becomes more constrained. Neutrino masses are generated through the type-I seesaw mechanism. As in the charged lepton sector, we assume that the couplings within each neutrino mass matrix have equal magnitudes, but with different relative phases.
 
\subsection{Charged lepton masses}

The charged lepton sector plays a crucial role in determining the value of the modulus $\tau$ within the present framework. Since the couplings are assumed to be universal, the structure of the charged lepton mass matrix is entirely controlled by the modular weights of right handed charged leptons and the value of $\tau$. As a result, reproducing the observed charged lepton mass hierarchy provides a direct constraint on the allowed regions of the modulus.

In this work, we systematically search for values of $\tau$ for which the experimentally measured charged lepton masses can be obtained. This analysis is performed under the assumption of universal modular invariant couplings, ensuring that the hierarchy is not introduced through arbitrary parameters but instead emerges from the modular structure.

The charged leptons are assigned to singlet representations of the $A_4$ symmetry, while their corresponding modular weights are denoted as in Table~(\ref{assignment1}). These assignments determine the transformation properties of the fields and the structure of the allowed Yukawa interactions. Consequently, the charged lepton Lagrangian takes the general form
\bea
\cal{L}&=&\lambda_1 E_1^{(k_{E_1})} H (L^{(k_L)} \otimes Y_3^{(k_1)}(\tau) )_1 +\lambda_2 E_2^{(k_{E_2})} H(L^{(k_L)} \otimes Y_3^{(k_2)}(\tau))_1^{\prime}\nonumber\\&+&\lambda_3 E_3^{(k_{E_3})} H(L^{(k_L)} \otimes Y_3^{(k_3)}(\tau))_1^{\prime\prime},\label{chargedleptonLagrangian}
\eea
The modular weights must follow the modular invariance condition in Eq.(\ref{modular_condition_re})
\bea
k_L+k_{H}+k_{E_1}&=&k_1,\nonumber\\
k_L+k_{H}+k_{E_2}&=&k_2,\nonumber\\
k_L+k_{H}+k_{E_3}&=&k_3,
\eea
where $k_{Ei}$ denote the modular weights of the right handed charged leptons and $k_L$ is the modular weight of the lepton doublet.
\begin{center}
\begin{table}
\begin{tabular}{|c|c|c|c|c|c|c|}
  \hline
  fields & L & $E_1$ &$E_2$ &$E_3$& $N$ &$ H $  \\
   \hline
  $A_4$ & 3 & 1 & $1^{\prime\prime}$ &$1^{\prime}$ & 3 & 1 \\
   \hline
  $k_I$ & 1 & $k_{E_1}$ & $k_{E_2}$ & $k_{E_3}$ & $k_N$ & 0  \\
  \hline
  \end{tabular}
\caption{The assignments of flavors within the framework of $A_4$ and the corresponding modular weight $k_I$.}\label{assignment1}
\end{table}
 \end{center}
Imposing universal couplings, $\lambda_1=\lambda_2=\lambda_3=\lambda$, eliminates any source of hierarchy from the Yukawa coefficients. Consequently, the observed charged lepton mass hierarchy is generated entirely by the modular structure, with the modular weights of the right handed charged leptons playing the central role. 
  
The structure of the charged lepton mass matrix is given by:
\bea
M_e&=&\lambda~v \left(
      \begin{array}{ccc}
        Y_{3,1}^{(k_1)}(\tau)  & Y_{3,3}^{(k_1)}(\tau)  & Y_{3,2}^{(k_1)}(\tau)  \\
       Y_{3,2}^{(k_2)}(\tau)  & Y_{3,1}^{(k_2)} (\tau) & Y_{3,3}^{(k_2)}(\tau)  \\
       Y_{3,3}^{(k_3)}(\tau)  & Y_{3,2}^{(k_3)}(\tau)  & Y_{3,1}^{(k_3)} (\tau) \\
      \end{array}\right).
      \eea
Physical masses are obtained from the Hermitian matrix $H_e = M_e M_e^\dagger,$ whose eigenvalues determine the charged lepton masses up to an overall scale.
Diagonalizing $H_e$ yields
\begin{equation}
U_e^\dagger H_e U_e=
\text{diag}(m_e^2,m_\mu^2,m_\tau^2).
\end{equation}

We perform a systematic analysis of the charged lepton mass matrix by scanning over all allowed assignments of the right handed charged lepton modular weights $k_{E_i}$, constrained by the requirement that the modular form weights satisfy $k_i \in \{-4,-2,\dots,6\}$. For definiteness, we fix the modular weights of the lepton doublet and Higgs field to $k_L=1$ and $k_H=0$, respectively. Under these assumptions, the right handed weights $k_{E_i}$ are scanned over the odd integer range $[-5,5]$. Therefore, the charged lepton mass matrix is determined by the modulus $\tau$ and the charged lepton weights $k_{E_i}$, namely $M_e = M_e(\tau, k_{E_i})$. 

An important result of this scan is that the charged lepton mass hierarchy can be reproduced at specific values of $\tau$, located around modular fixed points, with specific choices of modular weight assignments. Consequently, the charged lepton hierarchy depends only on the modular structure.
Particularly, we explore regions in the vicinity of the fixed points $i$, $\pm(0.5+i\,0.5)$, and $\pm\left(\frac{1}{2}+i\frac{\sqrt{3}}{2}\right)$, and identify three distinct classes of solutions:

\begin{enumerate}
\item \textbf{Solution I:} \\
This solution is realized for
\begin{equation}
\tau_{1,2} = \pm\left(0.497672877 + i~0.5434468\right),
\end{equation}
with modular weight assignments
\begin{equation}
(k_{E1},k_{E2},k_{E3}) =
(3,-3,-5),~
(-3,-5,3),~
(-5,3,-3).
\end{equation}
The overall factor, $\lambda v$, is fixed to match the recent experimentally charged lepton masses in \cite{pdg}.
The above choices of modular weights reproduce the charged lepton masses with very high precision,
\begin{equation}
(m_\tau, m_\mu, m_e) =
(1.77693,~0.105658,~0.000510998)\ \text{GeV},
\end{equation}
for $\lambda v=3.10124$ GeV.
\item \textbf{Solution II:} \\
A second class of solutions arises close to the fixed point $i$,
\begin{equation}
\tau_{3,4} = \pm\left(0.0479066201 + i~0.9989482068\right),
\end{equation}
with modular weights
\begin{equation}
(k_{E1},k_{E2},k_{E3}) =
(-5,-1,-1),~
(-1,-5,-1),~
(-1,-1,-5).
\end{equation}
After adjusting the overall factor, $\lambda v$, this configuration also reproduces the charged lepton masses at a high level of accuracy,
\begin{equation}
(m_\tau, m_\mu, m_e) =
(1.77693,~0.105658,~0.000510998)\ \text{GeV},
\end{equation}
at $\lambda v=0.358825$ GeV.
\item \textbf{Solution III:} \\
A third solution is found near the elliptic region,
\begin{equation}
\tau_{5,6} = \pm\left(0.5 + i~0.8657828\right),
\end{equation}
with modular weights
\begin{equation}
(k_{E1},k_{E2},k_{E3}) =
(-5,1,5),~
(1,5,-5),~
(5,-5,1).
\end{equation}
The resulting masses are
\begin{equation}
(m_\tau, m_\mu, m_e) =
(1.76713,~0.107337,~0.000528044)\ \text{GeV},
\end{equation} for $\lambda v=0.549451$ GeV.
It is clear that this point reproduce the hierarchical structure with small deviations at the percent level.
\end{enumerate}

\begin{table}[t]
\centering
\begin{tabular}{c c c c}
\hline
Solution & $\tau$ & $(k_{E1},k_{E2},k_{E3})$ & Mass prediction (GeV) \\
\hline
I &
$\pm\left(0.497672877 + i~0.5434468\right)$ &
$(3,-3,-5)$ permutations &
$(1.77693,~0.105658,~0.000510998)$ \\

II &
$\pm\left(0.0479066201 + i~0.9989482068\right),$ &
$(-5,-1,-1)$ permutations &
$(1.77693,~0.105658,~0.000510998)$ \\

III &
$\pm(0.5 + i~0.8657828)$ &
$(-5,1,5)$ permutations &
$(1.76713,~0.107337,~0.000528044)$ \\
\hline
\end{tabular}
\caption{Values of the modulus $\tau$ and modular-weight assignments reproducing the charged lepton masses with universal couplings.}
\label{tab:charged_lepton_tau}
\end{table}

The viable solutions are listed briefly in Table~\ref{tab:charged_lepton_tau} for ease of reference. It is noticed that the solutions linked to $\tau_{5,6}$ show significant differences from the experimental values. Consequently, in the following neutrino sector analysis, we consider the solutions concerning $\tau_{1,2}$ and $\tau_{3,4}$. These results indicate that the charged lepton mass hierarchy exerts strong constraints on $\tau$, restricting it to lie in the neighborhood of particular points in the modular domain. More specifically, the feasible solutions occur near the fixed points indicating that the hierarchy shown is induced by small deviations from residual symmetries. 

Within this framework the universal couplings assumption implies that the hierarchy is generated by the modular weights and the geometry of $\tau$. Therefore, the model does not depend on extra flavon fields, fine-tuning of parameters or ad hoc hierarchies for Yukawa couplings. The presence of multiple discrete modular weight solutions reveals that the same hierarchical organization can be realized in different ways within the modular framework. These charged lepton solutions are considered as inputs for the neutrino sector which in effect fixes the modulus $\tau$ value which reproduces charged lepton masses.

\subsection{Neutrino sector}
In order to generate neutrino masses, we adopt the type-I seesaw mechanism within the present framework. To preserve the minimal character of the model, we assume that the couplings entering each mass matrix are universal in magnitude, while allowing for independent complex phases. This assumption significantly reduces the number of free parameters and enhances the predictive power of the setup.

We extend the Standard Model by introducing three right handed neutrinos  $N$, which are transformed as a triplet under the $A_4$ symmetry with modular weight $k_N$. With these assignments, one can write down the most general set of modular invariant terms consistent with the symmetry for arbitrary values of $k_N$.
The resulting modular $A_4$ invariant Lagrangian can be written in the general form
\bea
\cal{L}&=&g_1~(N^{(k_N)} H L^{(k_L)})_{3S}Y_3^{(k_4)}(\tau)+g_2(N^{(k_N)} H L^{(k_L)}))_{3A}Y_3^{(k_4)}(\tau)+g_3
(N^{(k_N)} H L^{(k_L)})_1 Y_1^{(k_4)}(\tau)\nonumber\\&+&g_4
(N^{(k_N)} H L^{(k_L)})_1^{\prime\prime} ~Y_{1^{\prime}}^{(k_4)}(\tau)+\Lambda \Big(f_1 (N^{(k_N)}  N^{(k_N)})_{3S}Y_3^{(k_5)}(\tau)\nonumber+f_2 (N^{(k_N)}  N^{(k_N)})_1 Y_1^{(k_5)}(\tau)\\&+&f_3 (N^{(k_N)}  N^{(k_N)})_1^{\prime\prime}Y_{1^{\prime}}^{(k_5)}(\tau)\Big),\label{Lagrangian}
\eea
The parameter $\Lambda$ represents the scale of the non renormalizable operators and sets the overall mass scale of the right handed neutrinos, thereby determining the characteristic scale of the seesaw mechanism.

The coupling constants $g_1$ and $f_1$ are associated with the symmetric triplet contributions arising from the tensor product of two $A_4$ triplets. In contrast, $g_2$ corresponds to the antisymmetric triplet component. The coefficients $g_3$ and $f_2$ are linked to singlet modular forms of the type $Y_1^{(k_{4,5})}$, while $g_4$ and $f_3$ multiply the nontrivial singlet $Y_{1'}^{(k_{4,5})}$.

It is important to note that the modular form $Y_{1'}^{(k_{4,5})}$ is non vanishing only for the specific case $k_i=4$. Consequently, the corresponding terms contribute only when this condition is satisfied, which restricts the structure of the mass matrices for other values of the modular weights.

Imposing modular invariance, as expressed in Eq.(\ref{modular_condition_re}), leads to specific relations among the modular weights of the fields and modular forms, which must be satisfied to ensure consistency of the construction,
  \bea
  k_N+k_L&=&k_4,\nonumber\\
  2~k_N&=&k_5.
 \label{modular_invariance_nu} \eea  
The Dirac and Majorana mass matrices are of the form
\bea
M_D&=&v\left(
      \begin{array}{ccc}
        2g_1Y_{3,1}^{(k_4)}+g_3Y_1^{(k_4)} &(-g_1+g_2)Y_{3,3}^{(k_4)}   & (-g_1-g_2)Y_{3,2}^{(k_4)}+g_4Y_1^{\prime^{(k_4)}} \\
      (-g_1-g_2)Y_{3,3}^{(k_4)}  &2g_1Y_{3,2}^{(k_4)}+g_4Y_1^{\prime^{(k_4)}} & (-g_1+g_2)Y_{3,1}^{(k_4)} +g_3Y_1^{(k_4)} \\
        (-g_1+g_2)Y_{3,2}^{(k_4)}+g_4Y_1^{\prime^{(k_4)}} & (-g_1-g_2)Y_{3,1}^{(k_4)} +g_3Y_1^{(k_4)} &  2g_1Y_{3,3}^{(k_4)}\\
      \end{array}\right),\nonumber\\
M_R&=&\Lambda\left(
      \begin{array}{ccc}
        2f_1Y_{3,1}^{(k_5)}+f_2Y_1^{(k_5)} &-f_1Y_{3,3}^{(k_5)}   & -f_1Y_{3,2}^{(k_5)}+f_3~Y_1^{\prime^{(k_5)}} \\
      -f_1~Y_{3,3}^{(k_5)}  &2f_1Y_{3,2}^{(k_5)}+f_3~Y_1^{\prime^{(k_5)}} & -f_1~Y_{3,1}^{(k_5)} +f_2Y_1^{(k_5)} \\
        -f_1~Y_{3,2}^{(k_5)}+f_3~Y_1^{\prime^{(k_5)}} & -f_1~Y_{3,1}^{(k_5)} +f_2~Y_1^{(k_5)} &  2f_1~Y_{3,3}^{(k_5)}\\
      \end{array}\right).      
      \eea
In general, the couplings appearing in both the Dirac and Majorana mass matrices are complex parameters. In order to preserve the minimality of the model and avoid introducing unnecessary degrees of freedom, we consider that all couplings within a given matrix have the amplitude, while allowing for independent relative phases. This assumption significantly reduces the dimensionality of the parameter space.

Under this parametrization, the Dirac sector couplings can be expressed as
\begin{equation}
g_1 = g, ~~ g_2 = g\,e^{i\alpha}, ~~ g_3 = g\,e^{i\beta}, ~~ g_4 = g\,e^{i\gamma_D},
\end{equation}
while the Majorana sector couplings take the form
\begin{equation}
f_1 = f, ~~ f_2 = f\,e^{i\phi}, ~~f_3 = f\,e^{i\gamma_R}.
\end{equation}
As a result, the number of independent phases is reduced to three in the Dirac mass matrix and two in the right handed neutrino mass matrix. 

It is important to note that the modular form $Y_{1'}^{(k_{4,5})}$ vanishes unless the modular weights satisfy $k_{4,5}=4$. Therefore, for $k_{4,5} \neq 4$, the terms proportional to $g_4$ and $f_3$ are absent, reducing the number of physical phases to three.

The effective light neutrino mass matrix is then generated through the type-I seesaw mechanism,
\begin{equation}
M_\nu = - M_D M_R^{-1} M_D^T.
\end{equation}
To extract the physical neutrino masses and mixing parameters, we construct the Hermitian matrix
\begin{equation}
H_\nu = M_\nu M_\nu^\dagger,
\end{equation}
which is diagonalized by a unitary transformation,
\begin{equation}
U_\nu^\dagger H_\nu U_\nu = \mathrm{diag}(m_1^2, m_2^2, m_3^2).
\end{equation}
Finally, the leptonic mixing matrix is obtained as
\begin{equation}
U_{\rm PMNS} = U_e^\dagger U_\nu,
\end{equation}
where $U_e$ is the unitary matrix that diagonalizes the charged lepton sector.

\subsection{Charged lepton permutation selection}

In flavor model building, an intrinsic ambiguity exists from the ordering of the charged lepton eigenvectors. Since the leptonic mixing matrix is constructed as $U_{\rm PMNS} = U_e^\dagger U_\nu$, any permutation of the eigenvectors of the Hermitian matrix $H_e = M_e M_e^\dagger$ effectively corresponds to a different embedding of the charged lepton sector relative to the neutrino sector. 

In fact, this ambiguity is resolved in the conventional approach by fixing a specific basis or by imposing additional residual symmetries by hand. Yet, for charged lepton and neutrino sectors originating from distinct symmetry breaking patterns, the relative alignment between the two sectors acquires physical significance~\cite{Goswami:2025jde, Hagedorn:2017zks,Altarelli:2010gt,King:2013eh,Feruglio:2012cw,Holthausen:2012dk}. 
The two  sectors are governed by the modulus $\tau$ and the associated modular weights in modular flavor models. For this reason the alignment between $U_e$ and $U_\nu$ is not arbitrary but is dynamically determined by the underlying modular structure. That is, different permutations of the charged lepton eigenvectors can lead to genuinely distinct phenomenological predictions~\cite{Kobayashi:2018scp,Criado:2018thu,King:2020qaj}. 

In the current paper we systematically study all six possible permutations of these charged lepton eigenstates. 
For completeness, we label the six permutations as
\begin{align}
\text{permIndex }1 &: (e,\mu,\tau), ~~~~
\text{permIndex }2 : (e,\tau,\mu), \nonumber \\
\text{permIndex }3 &: (\mu,e,\tau), ~~~~
\text{permIndex }4 : (\mu,\tau,e), \nonumber \\
\text{permIndex }5 &: (\tau,e,\mu), ~~~~
\text{permIndex }6 : (\tau,\mu,e).
\label{permutation}
\end{align}

\section{Numerical Analysis}\label{sec:results}

The structure of the effective neutrino mass matrix depends on the modulus $\tau$, the modular weight of the right handed neutrinos $k_N$, and the set of phase parameters.

In this section, we examine the phenomenological viability of the model through a comprehensive numerical scan over the relevant parameter space.

The modulus $\tau$ is fixed by the charged lepton sector to the values
\begin{equation}
\tau_{1,2}= \pm(0.49767256+0.543455\,i), \qquad
\tau_{3,4}=\pm(0.0479085+0.9989481165\,i),
\label{selected_tau}
\end{equation}
which successfully reproduce the observed charged lepton mass hierarchy under the assumption of universal couplings.

The modular weights $k_4$ and $k_5$ are constrained to be even. Therefore, from the modular invariance condition in Eq.(\ref{modular_invariance_nu}), nontrivial modular forms arise simultaneously in the Dirac and Majorana sectors only for discrete odd values of the right handed neutrino weight,
$
k_N = -1,~ 1,~ 3,\label{k_N}
$
with the corresponding modular weights satisfying $k_{4,5}\in[-2,6]$. Accordingly, our scan is restricted to these discrete values of $k_N$.

The number of independent physical phases is determined by the allowed modular weights as follows:
\begin{enumerate}

\item For $k_N=-1, ~1$, the modular weights in Majorana sector, $k_{4,5} \neq 4$, so the modular form $Y_{1'}^{4,5}$ vanishes, and the corresponding terms in the Lagrangian are absent. In this case, the number of the independent phases reduced to three,
 i.e.
$
M_{\nu} = M_{\nu}(\tau, \alpha, \beta, \phi, k_N).
$
\item In the right handed neutrino sector, the contribution from $Y_{1'}^{k_5}$ is always absent, since the condition $k_{5}=2k_N$ together with the allowed odd values of $k_N$ does not permit $k_5=4$. Consequently, the phase $\gamma_R$ is not physical in this setup.

\item For $k_N=3$, the Dirac sector allows $k_4 = k_N + k_L = 4$, so that the modular form $Y_{1'}^{(4)}$ contributes. In this case, an additional phase $\gamma_D$ becomes relevant, and the total number of independent phases increases to four.
\end{enumerate}
\begin{table}[t]
\begin{center}
\begin{tabular}{|c|c|c|c|c|c|c|c|}
  \hline
  &$\frac{\Delta m^2_{12}}{(10^{-5}~\text{eV}^2)}$ & $\frac{\Delta m^2_{31}}{(10^{-3}~\text{eV}^2)}$ & $r=\frac{\Delta m^2_{12}}{\Delta m^2_{31}}$&$\theta_{12}/^{\circ}$ & $\theta_{23}/^{\circ}$ &$ \theta_{13}/^{\circ} $ \\
   \hline
  best fit&7.39&2.525&0.0292&33.82&49.6&8.61\\
   \hline
  $3\sigma$ range& 6.79-8.01 & 2.41-2.611 &0.0258-0.033& 31.61-36.27 & 40.3-52.4& 8.22-9  \\
  \hline
\end{tabular}
\end{center}
\caption{The 3$\sigma$ range for neutrino mixing parameters and the squares of mass differences, as presented in \cite{Esteban:2018azc}, pertains to the scenario of normal hierarchy (NO).}\label{best_fit}
\end{table}

For each value of $\tau$ in Eq.~(\ref{selected_tau}), we perform a scan over the full range of phases, the allowed values of $k_N$, and all possible charged lepton permutations entering the PMNS matrix. The viable points must satisfy the $3\sigma$ experimental constraints on the mixing angles $(\theta_{12},\theta_{13},\theta_{23})$ and the mass squared ratio
$
r = \frac{\Delta m_{21}^2}{\Delta m_{31}^2},
$
as summarized in Table~\ref{best_fit} for the case of normal ordering.

 To calculate physical quantities such as the effective Majorana mass $m_{ee}$ and the total neutrino mass $\Sigma m_i$ for each accepted point, we fix the overall factor $\kappa = \frac{v^2 g^2}{\Lambda f}$ by matching the calculated mass squared difference to its experimental value. For that reason, we define
\begin{equation}
\kappa=\sqrt{\frac{\Delta m_{31,\mathrm{exp}}^2}{m^2_3-m_1^2}},
\end{equation}
where $m_1$ and $m_3$ are unscaled masses obtained from the numerical diagonalization, and $\Delta m_{31,\mathrm{exp}}^2$ is the experimental mass square difference. The physical neutrino masses are then given by
\begin{equation}
m_i^{\mathrm{phys}}=\kappa\, m_i,
\end{equation}
so that the observables
\begin{equation}
m_{ee}=\left|\sum_{i=1}^3 U_{ei}^2 m_i^{\mathrm{phys}}\right|,
\qquad
\Sigma m_i = \sum_{i=1}^3 m_i^{\mathrm{phys}},
\end{equation}
can be evaluated in physical units.

One key conclusion from numerical analysis results is that solutions are only found for
$
\tau_{1,2}= \pm(0.49767256+0.543455\,i),
$
together with all three charged lepton modular weight assignments
\begin{equation}
(k_{E1},k_{E2},k_{E3}) = (-3,-5,3),\quad (-5,3,-3),\quad (3,-3,-5).
\end{equation}
These solutions exist only for normal neutrino mass ordering and uniquely select the right handed neutrino modular weight
\begin{equation}
k_N = -1.
\end{equation}

As a consequence, the parameter space is reduced to three independent phases $(\alpha,\beta,\phi)$, and no viable points are found for other values of $k_N$. 
This result indicates a robust predictive capability of the framework: once $\tau$ is fixed by the charged lepton sector, the neutrino data select a unique modular weight configuration. Another nontrivial characteristic is that effective solutions exist only for a limited number of charged lepton permutations. It means that the model not only limits the continuous variables but also select a preferred relative alignment of the charged lepton to the neutrino sectors. In the following, we discuss results on each charged lepton modular weight assignment, in detail.

\subsection{Case I: $(k_{E1},k_{E2},k_{E3})=(-3,-5,3)$}
We start with the modular weight assignment $(k_{E1},k_{E2},k_{E3}) = (-3,-5,3)$. In this assignment, the phenomenologically viable solutions are highly constrained with respect to the charged lepton permutations. In particular, the numerical analysis indicates that only a single permutation, namely $\text{permIndex }1$ (see Eq.(\ref{permutation})), satisfies all experimental constraints. This behavior indicates that the model for this assignment strongly select a specific permutation, reflecting a nontrivial selection of the charged lepton alignment.

Since the couplings are taken to have equal magnitudes, the flavor structure is mainly controlled by the phase parameters and the modular weights. As a consequence, the allowed parameter space shows clear correlations among the phases. In particular, the solutions are confined to narrow and well defined regions in the $(\alpha,\beta)$, $(\alpha,\phi)$, and $(\beta,\phi)$ planes. This indicates that, the phases are correlated and don't vary independently when the neutrino observables are fixed.

\begin{figure}[t]
\centering
\includegraphics[width=0.48\linewidth]{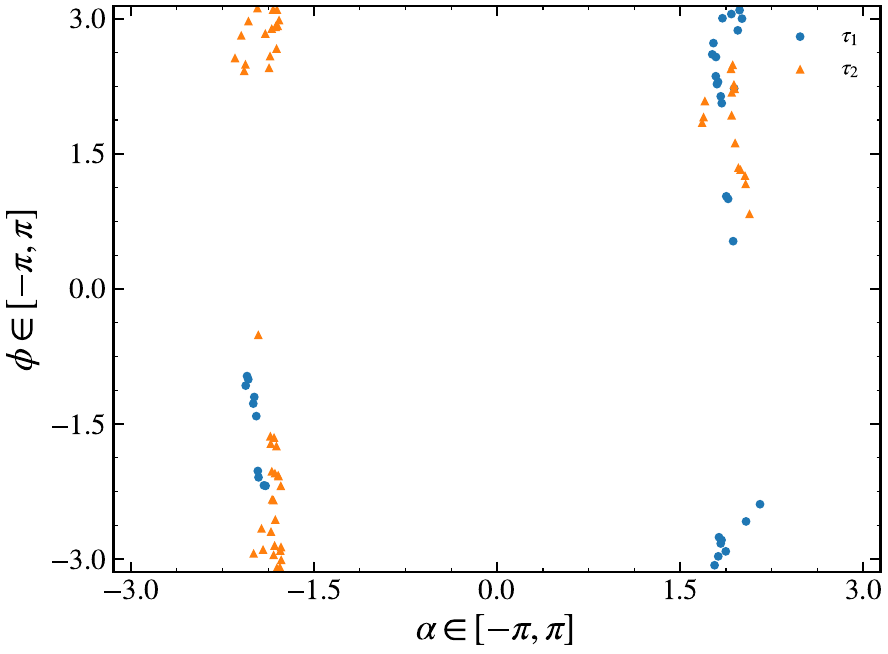}\includegraphics[width=0.48\linewidth]{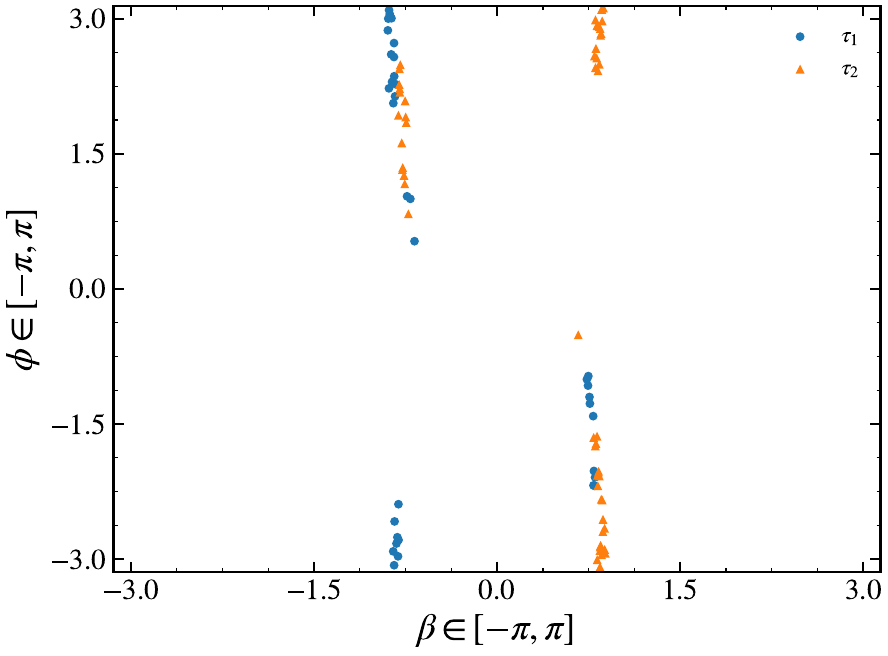}
\caption{The allowed values of the phases $\alpha, ~\beta$ and $\phi$ that yield to the observed data for NO with $k_N=-1$ and $k_E=(-3,-5,3)$. Circles correspond to $\tau_1$ and triangles to $\tau_2$.}
\label{fig:alpha_beta_phi_tau1_tau2_case1}
\end{figure}

These correlations appear in relatively compact regions, suggesting that the limited charged lepton alignments leads to a more constrained phase structure in the neutrino sector. Representative examples of these phase correlations are shown in Fig.~\ref{fig:alpha_beta_phi_tau1_tau2_case1}.

\begin{figure}[t]
\centering
\includegraphics[width=0.48\linewidth]{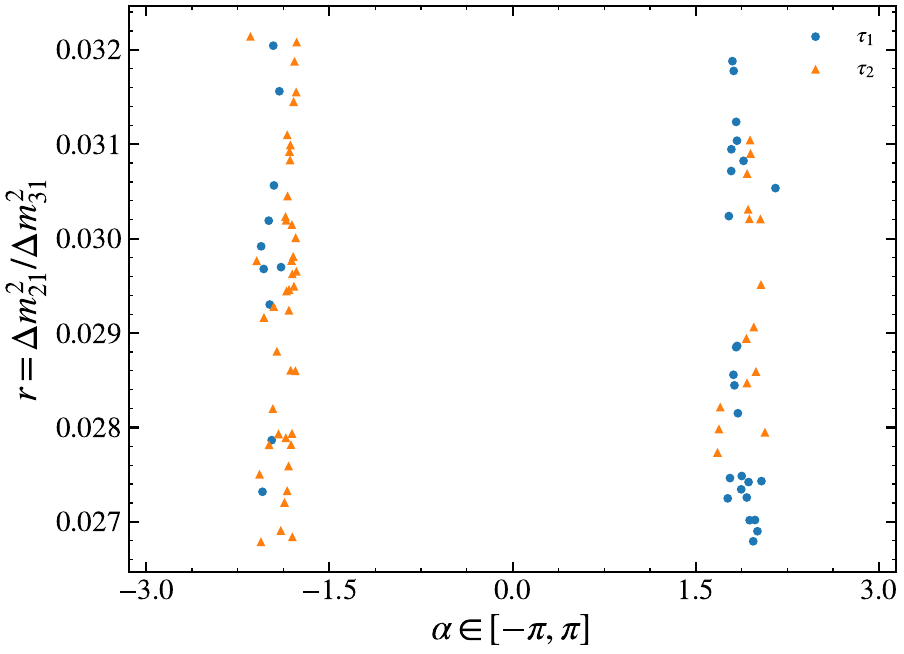}
\includegraphics[width=0.48\linewidth]{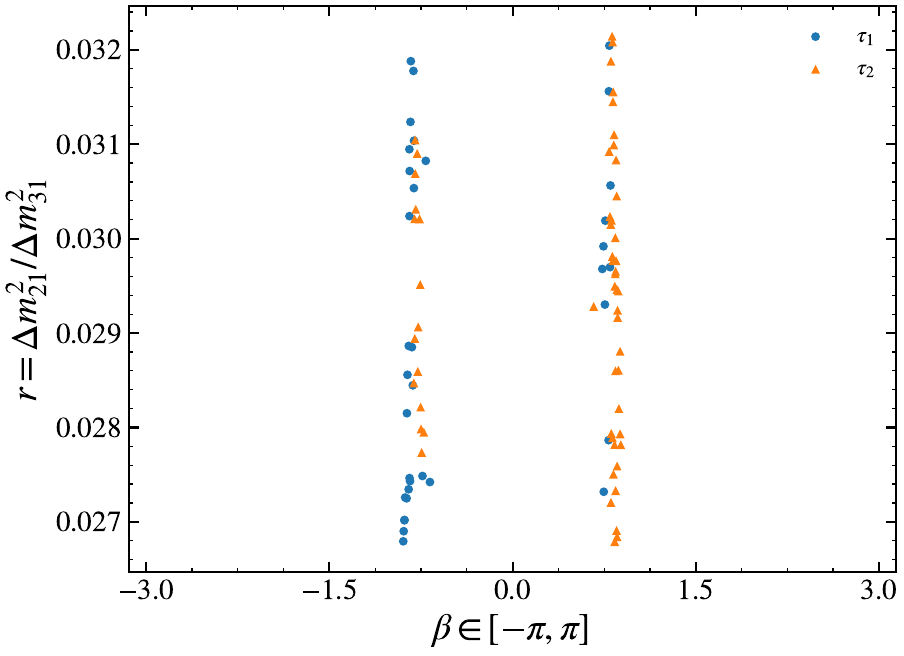}
\caption{Correlation between the ratio $r$ and the phases for NO with $k_N=-1$ and $k_E=(-3,-5,3)$. Circles correspond to $\tau_1$ and triangles to $\tau_2$.}
\label{fig:r_alpha_tau1_tau2_case1}
\end{figure}

To illustrate how the experimentally measured mass splitting ratio $r=\frac{\Delta m_{21}^2}{\Delta m_{31}^2}$ constrains the phase parameters, Fig.~\ref{fig:r_alpha_tau1_tau2_case1} displays the ratio $r$ as a function of $\alpha$ and $\beta$ as examples. The allowed points are confined to narrow bands in these parameter planes, clearly demonstrating that the ratio $r$ imposes strong constraints on the phase structure of the model.

\begin{figure}[t]
\centering
\includegraphics[width=0.48\linewidth]{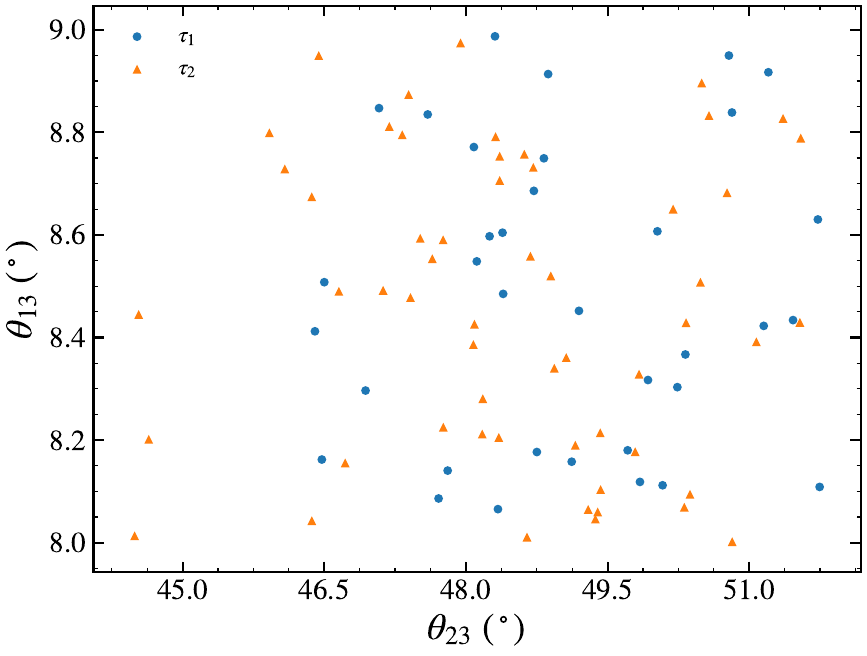}
\includegraphics[width=0.48\linewidth]{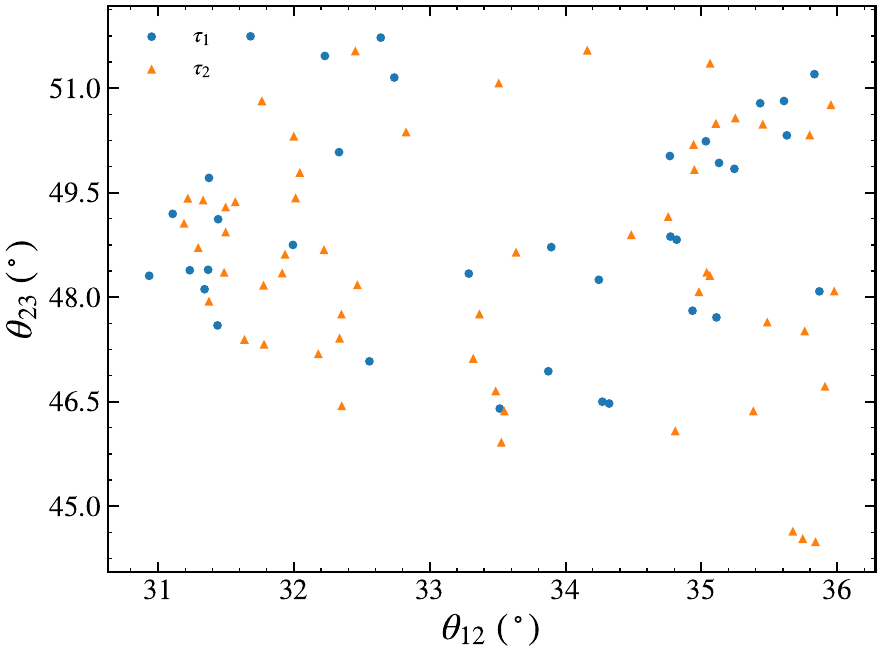}
\caption{Allowed $3\sigma$ regions of the mixing angles for NO with $k_N=-1$ and $k_E=(-3,-5,3)$. Circles correspond to $\tau_1$ and triangles to $\tau_2$.}
\label{fig:theta23_theta13_case1}
\end{figure}

The correlations among the mixing angles are similarly tight and highly predictive. The allowed points do not populate the full $3\sigma$ ranges uniformly, but instead form restricted and nontrivial patterns. These confirm that this charged lepton assignment is fully consistent with the observed PMNS structure for $k_N=-1$. These correlations are shown in Fig.~\ref{fig:theta23_theta13_case1}.

\begin{figure}[t]
\centering
\includegraphics[width=0.48\linewidth]{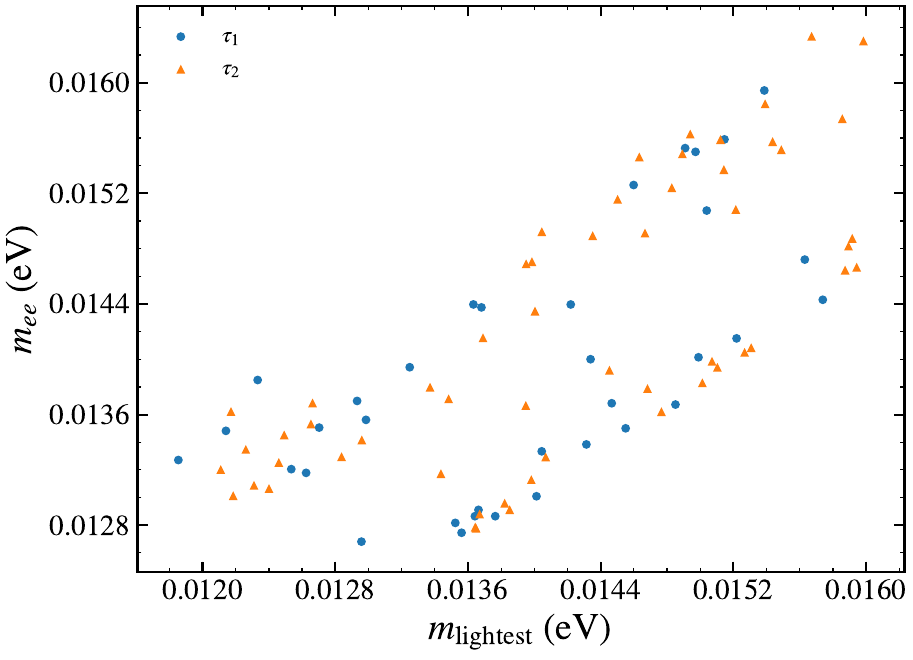}
\includegraphics[width=0.48\linewidth]{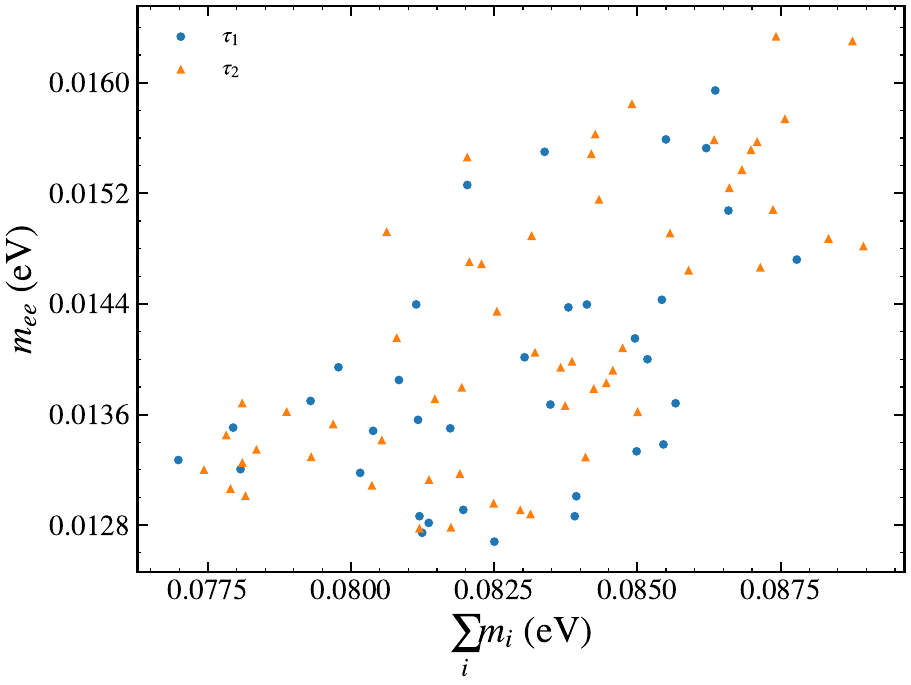}
\includegraphics[width=0.48\linewidth]{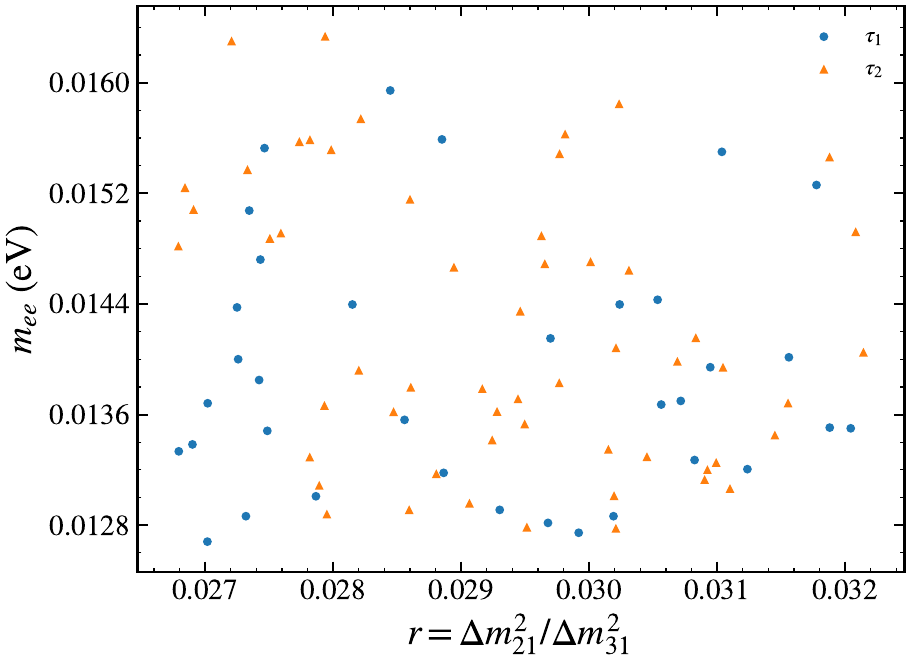}
\caption{Correlation between $m_{ee}$ and the lightest neutrino mass, the neutrino mass sum, and the mass-squared ratio $r$ for $k_E=(-3,-5,3)$.}
\label{fig:mee_case1}
\end{figure}

An important phenomenological aspect of the model is its prediction for the neutrinoless double beta decay parameter $m_{ee}$ and the cosmological neutrino mass sum $\Sigma m_i$. In Fig.~\ref{fig:mee_case1}, we show $m_{ee}$ as a function of the lightest neutrino mass $m_{\rm lightest}$. The allowed points are not randomly distributed but lie within a well defined correlated region, indicating that $m_{ee}$ becomes strongly constrained once the mixing angles and the mass square ratio $r$ are fixed.
The worthy feature is that a finite lower bound of $m_{ee}$ is obtained and the model does not allow it to become arbitrarily small within the viable region.

The correlation between $\Sigma m_i$ and $m_{ee}$ is shown in Fig.~\ref{fig:mee_case1}, where the viable points form a narrow band.  The reason is that cosmological limits on $\Sigma m_i$ convert directly into constraints on $m_{ee}$ within this model.
Finally, the last graph in Fig.~\ref{fig:mee_case1} illustrates the correlation between $m_{ee}$ and the ratio $r$. 

\subsection{Case II: $(k_{E1},k_{E2},k_{E3})=(-5,3,-3)$}
Next we  consider the modular weight assignment $(k_{E1},k_{E2},k_{E3}) = (-5,3,-3)$. As in the previous case, the solutions are found for both $\tau_1$ and $\tau_2$. In this case, the accepted points correspond to a single charged lepton permutation, $\text{permIndex }5$ (see Eq.~(\ref{permutation})), indicating that the model continues to favor a specific alignment. 

The allowed values of parameters that satisfy the experimental constraints are shifted compared to the $(-3,-5,3)$ case. This shows that the interplay between the charged lepton and neutrino sectors depends on the choice of modular weights, despite the charged lepton masses are reproduced equally in the two cases.

\begin{figure}[t]
\centering
\includegraphics[width=0.48\linewidth]{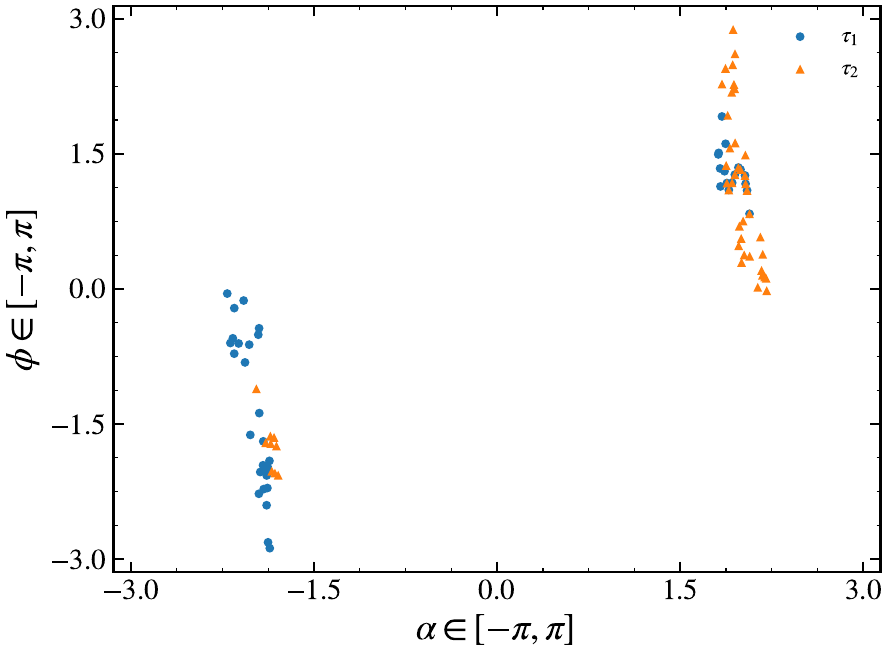}\includegraphics[width=0.48\linewidth]{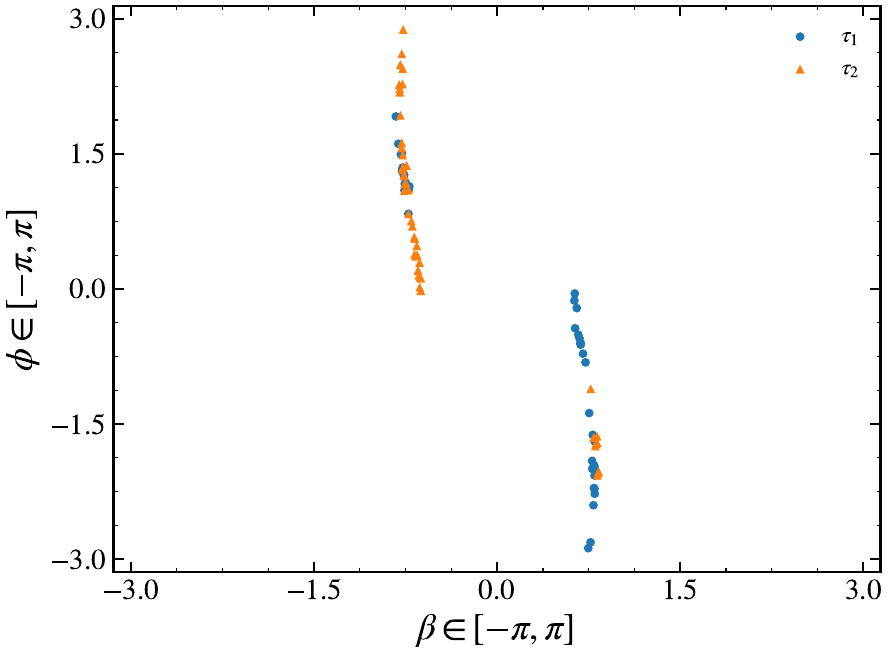}
\caption{The allowed values of the phases $\alpha, ~\beta$ and $\phi$ that yield to the observed data for NO with $k_N=-1$ and $k_E=(-5,3,-3)$. Circles correspond to $\tau_1$ and triangles to $\tau_2$.}
\label{fig:alpha_beta_phi_tau1_tau2_case2}
\end{figure}

As before, the phase correlations are still very organized. The possible solutions are limited to narrow regions in phase space. This means that they originate from specific phase alignments not  accidentally. The allowed phase ranges are shown in Fig.~\ref{fig:alpha_beta_phi_tau1_tau2_case2}.

\begin{figure}[t]
\centering
\includegraphics[width=0.48\linewidth]{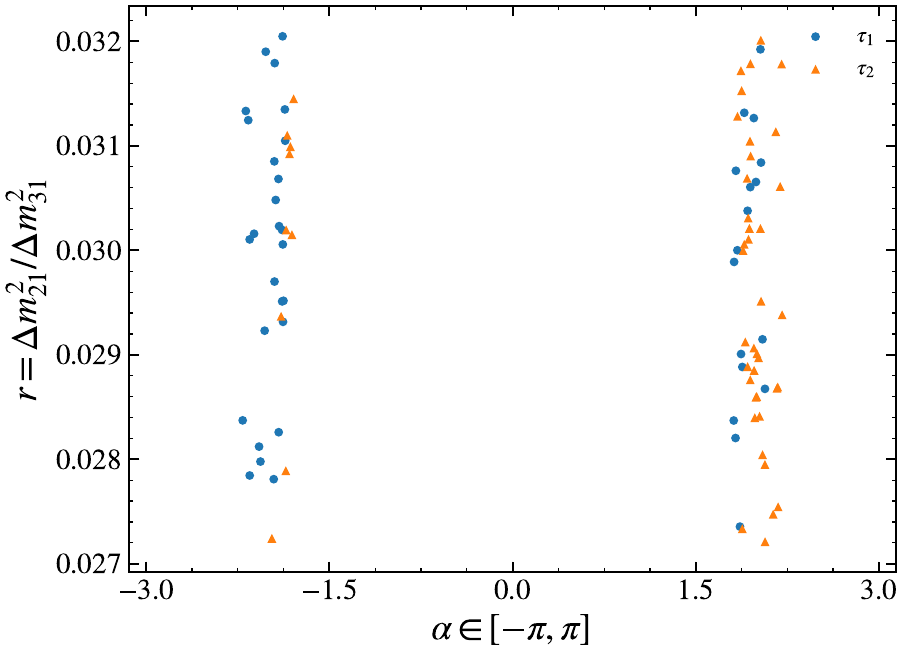}
\includegraphics[width=0.48\linewidth]{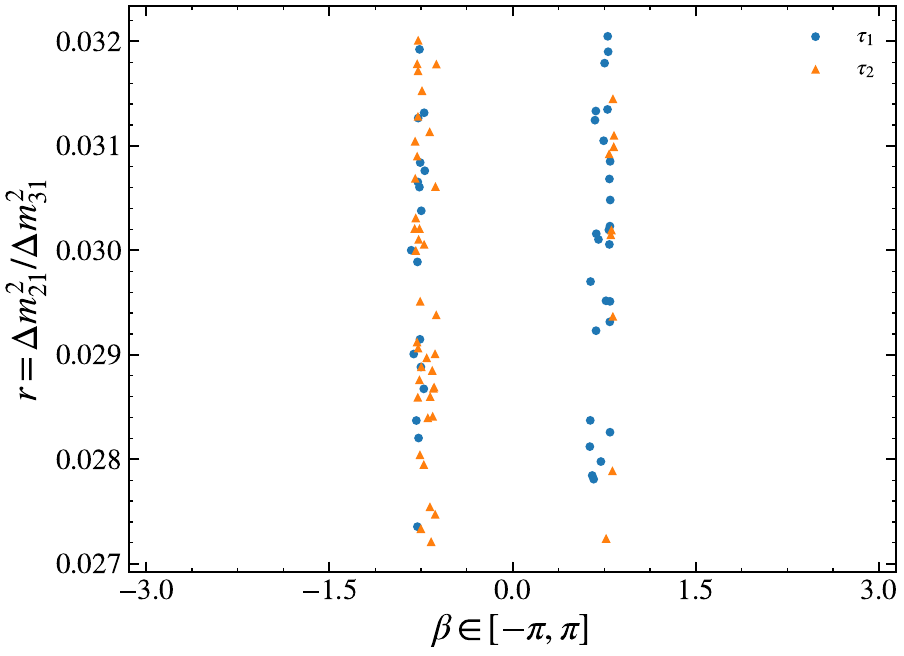}
\caption{Correlation between ratio $r$ and the phases for NO with $k_N=-1$ and $k_E=(-5,3,-3)$. Circles correspond to $\tau_1$ and triangles to $\tau_2$.}
\label{fig:r_alpha_tau1_tau2_case2}
\end{figure}

The correlations between the ratio $r$ and the phases are shown in (Fig. \ref{fig:r_alpha_tau1_tau2_case2}). These indicate that the measured neutrino mass splittings are connected to the phase structure that controls mixing. 

\begin{figure}[t]
\centering
\includegraphics[width=0.48\linewidth]{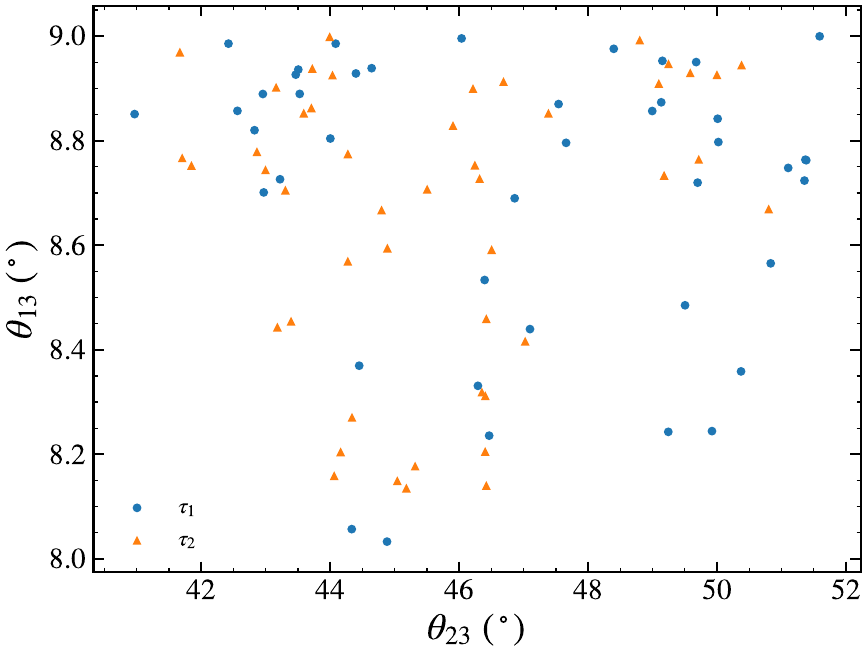}\includegraphics[width=0.48\linewidth]{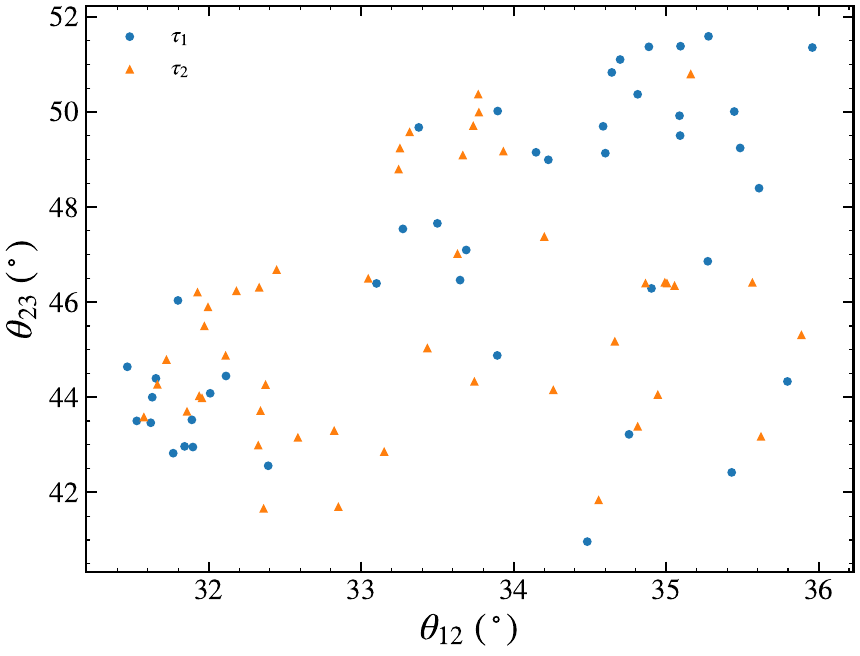}
\caption{Accepted points for NO with $k_N=-1$ and $k_E=(-5,3,-3)$. Circles correspond to $\tau_1$ and triangles to $\tau_2$.}
\label{fig:theta23_theta13_case2}
\end{figure}

The mixing angle correlations are similar in structure to those of the previous assignment, but the detailed distribution of accepted points is shifted. The mixing angle correlation is shown in Fig.~\ref{fig:theta23_theta13_case2}.

\begin{figure}[t]
\centering
\includegraphics[width=0.48\linewidth]{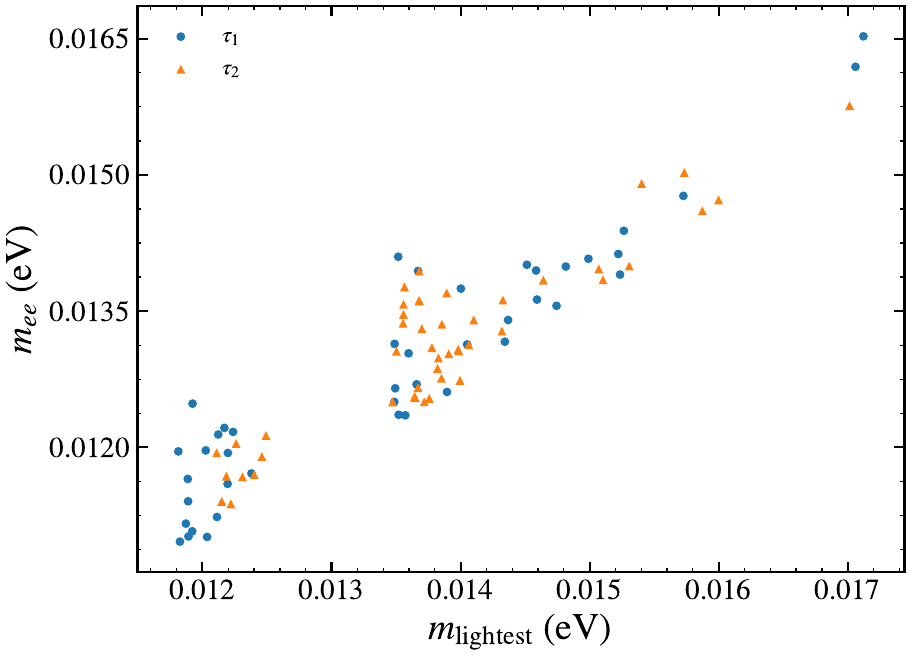}\includegraphics[width=0.48\linewidth]{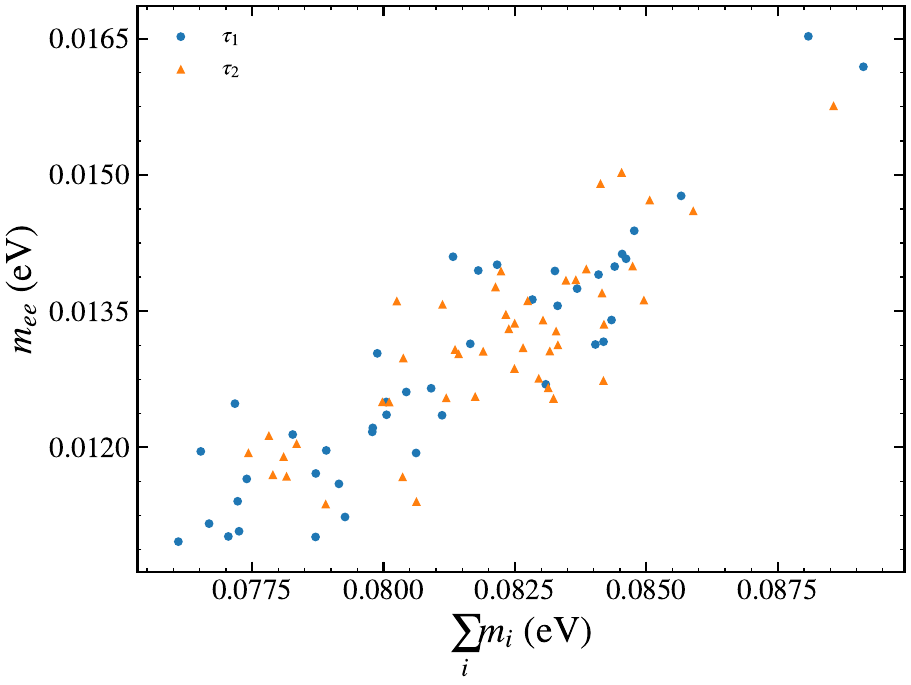}\\~\includegraphics[width=0.48\linewidth]{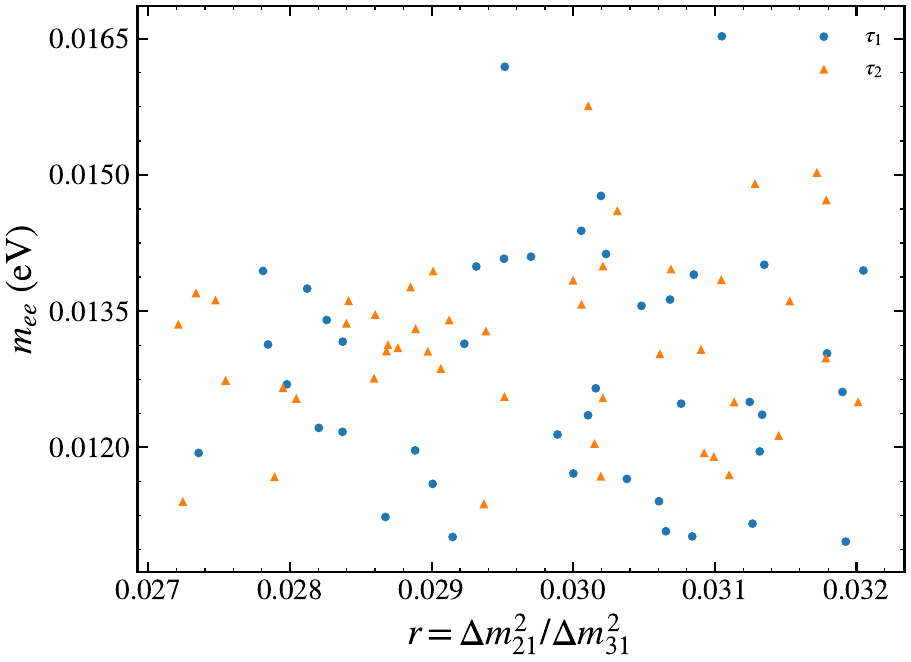}
\caption{Correlation between $m_{ee}$ and the lightest neutrino mass, the neutrino mass sum and mass square ratio $r$ for $k_E=(-5,3,-3)$.}
\label{fig:mee_case2}
\end{figure}

The neutrino mass observables also remain tightly constrained. The model predicts a limited region in the $(m_{ee},m_{\rm lightest})$ plane, a correlated band in the $(m_{ee},\Sigma)$ and $(m_{ee}, r)$ planes. 
The corresponding mass prediction is shown in Fig.~\ref{fig:mee_case2}.

\subsection{Case III: $(k_{E1},k_{E2},k_{E3})=(3,-3,-5)$}

\begin{figure}[t]
\centering
\includegraphics[width=0.48\linewidth]{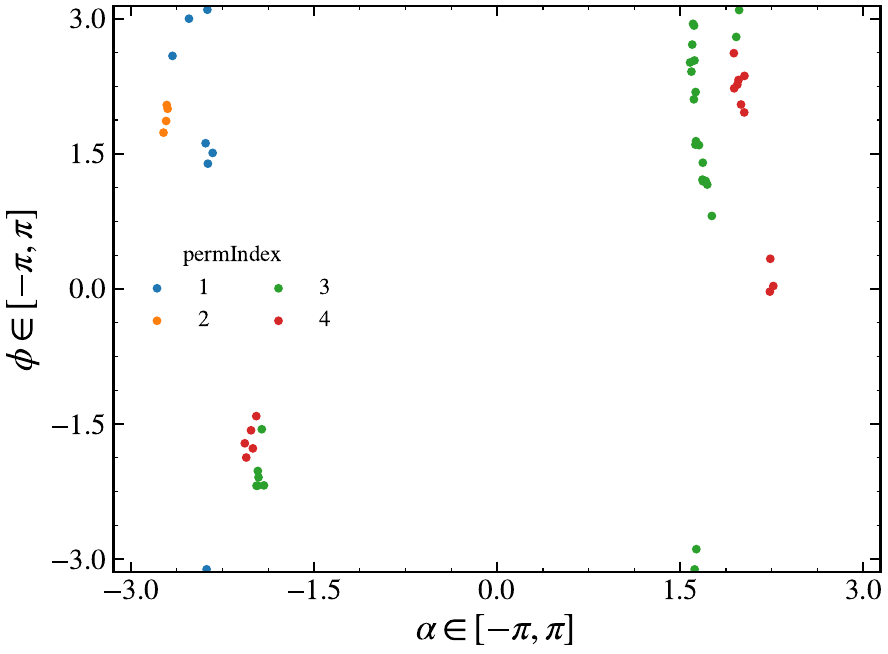}
\includegraphics[width=0.48\linewidth]{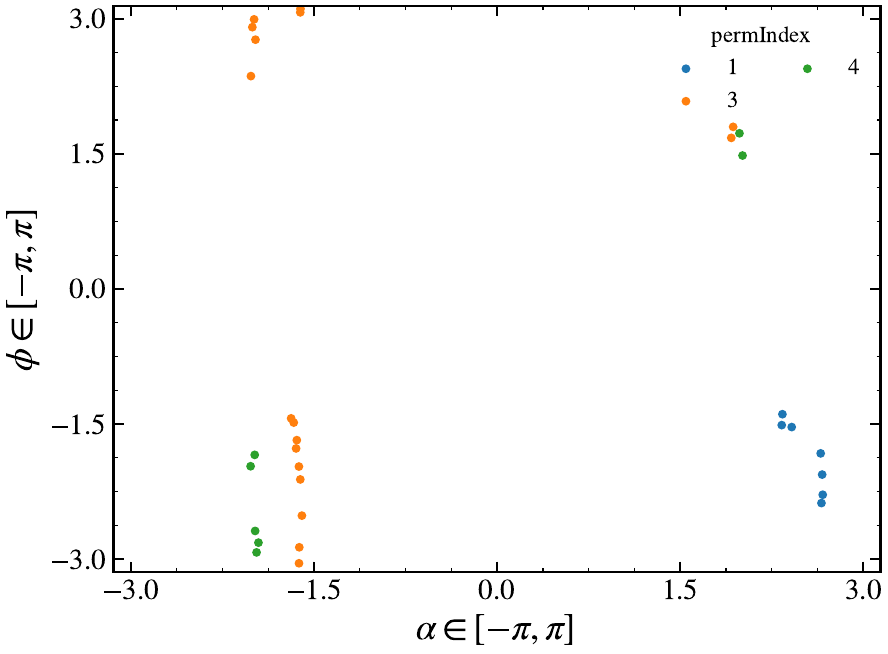}
\caption{Correlation between the phases $\alpha$ and $\phi$ for $\tau_1$ (left) and $\tau_2$ (right) for $k_E=(3,-3,-5)$. Colors indicate permutation index.}
\label{fig:alpha_phi}
\end{figure}

\begin{figure}[t]
\centering
\includegraphics[width=0.48\linewidth]{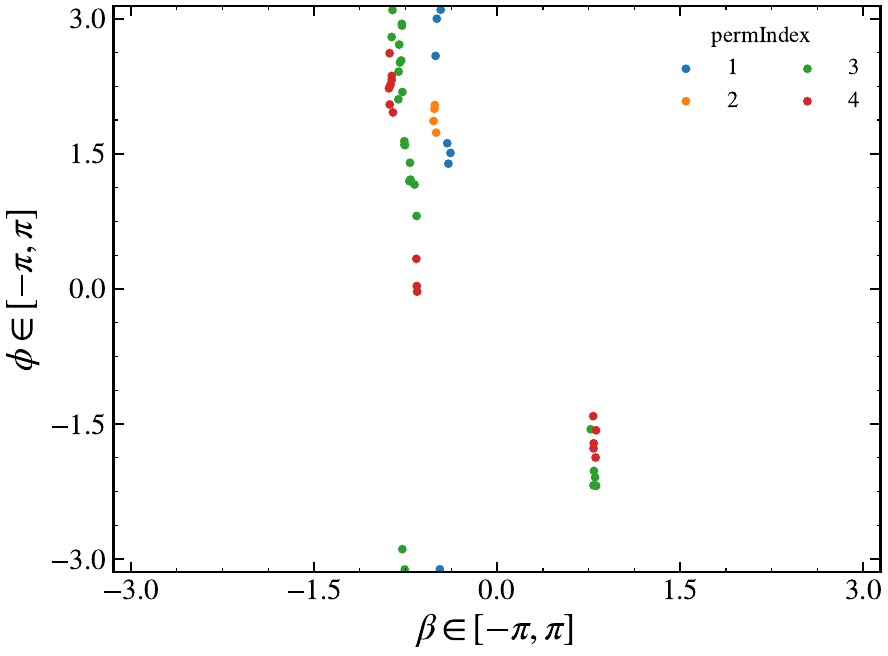}
\includegraphics[width=0.48\linewidth]{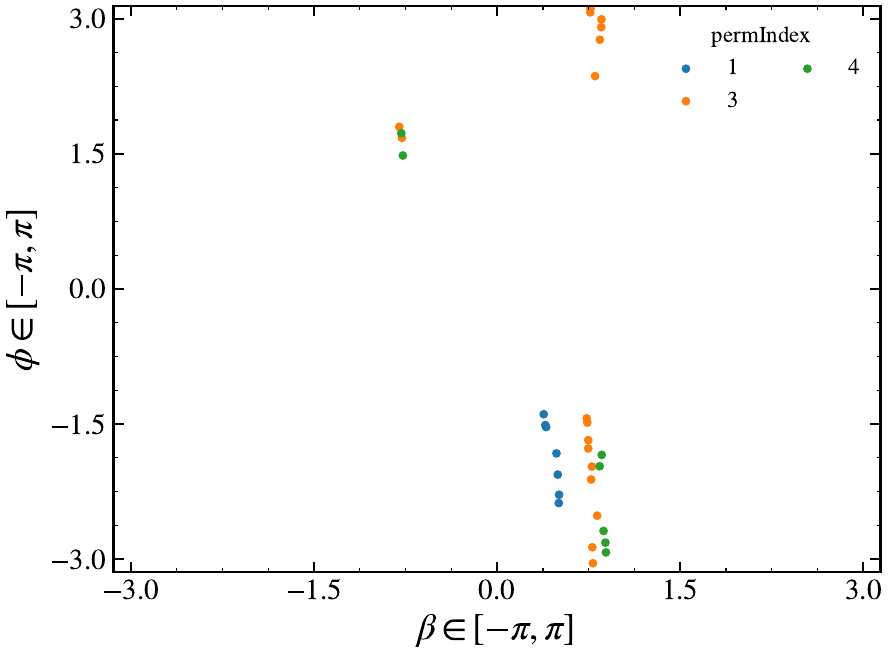}
\caption{Correlation between the phases $\beta$ and $\phi$ for $\tau_1$ (left) and $\tau_2$ (right) for $k_E=(3,-3,-5)$. Colors indicate permutation index.}
\label{fig:beta_phi}
\end{figure}

We finally consider the assignment
$
(k_{E1},k_{E2},k_{E3})=(3,-3,-5).
$
In this case, the results for $\tau_1$ and $\tau_2$ are presented separately, since more than one charged lepton permutation is allowed , see Figs.~\ref{fig:alpha_phi}-\ref{fig:beta_phi}. The different colors in these figures correspond to the viable permutation indices. Although the three charged lepton assignments lead to the same mass spectrum, their phenomenological implications differ when the neutrino sector is included. 

\begin{figure}[t]
\centering
\includegraphics[width=0.48\linewidth]{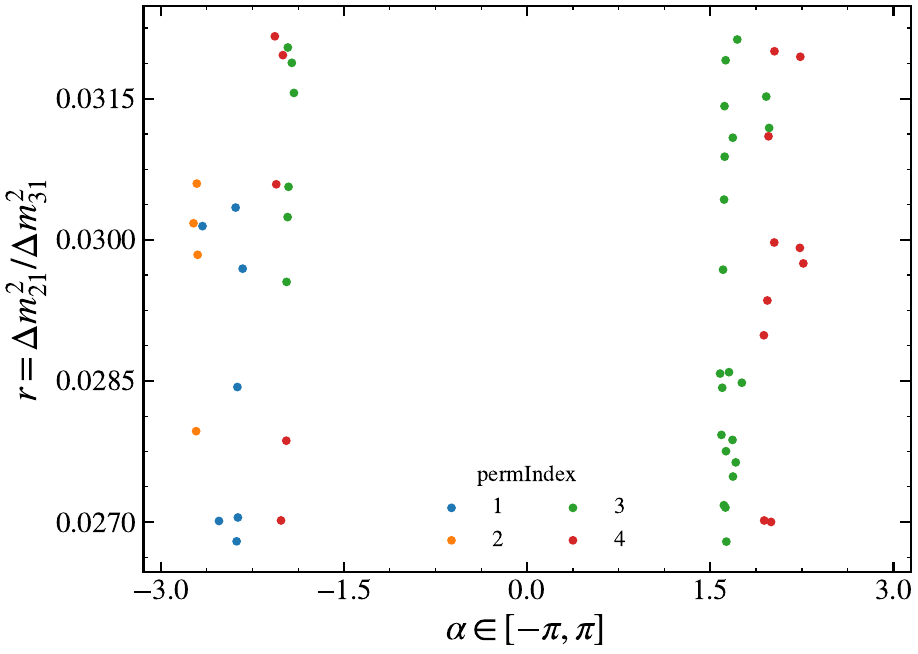}
\includegraphics[width=0.48\linewidth]{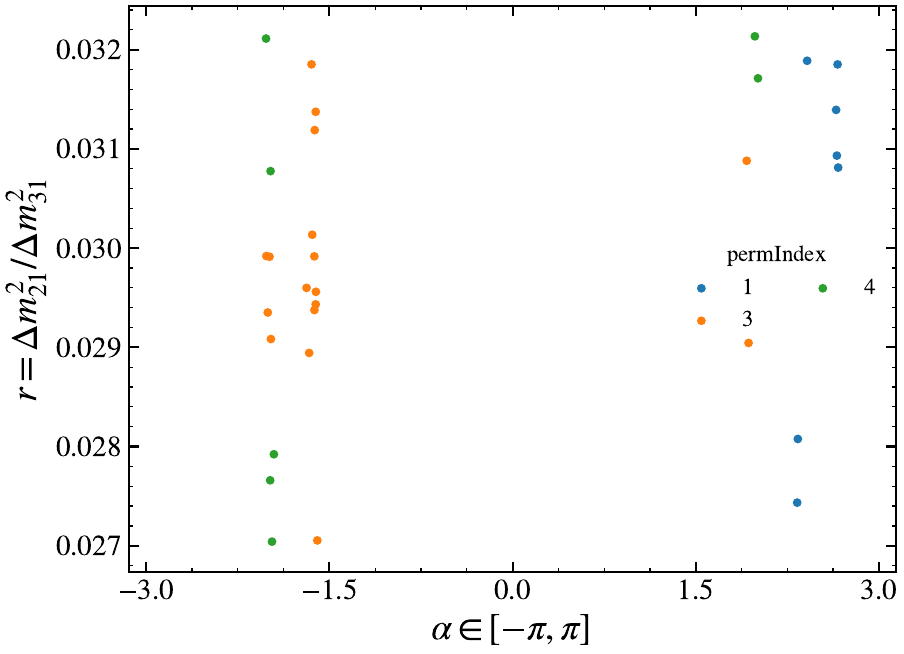}
\caption{Correlation between ratio $r$ and the phase $\alpha$ at NO with $k_N=-1$ and $k_E=(3,-3,-5)$ for $\tau_1$ (left) and $\tau_2$ (right). Colors indicate permutation index.}
\label{fig:r_alpha_tau1_tau2_case3}
\end{figure}

A representative phase correlation for $\tau_1$ and $\tau_2$ is shown in Fig.~\ref{fig:alpha_phi}-\ref{fig:beta_phi}. The correlations between the ratio $r$ and one of the phases are shown in Fig. \ref{fig:r_alpha_tau1_tau2_case3}.

\begin{figure}[t]
\centering
\includegraphics[width=0.48\linewidth]{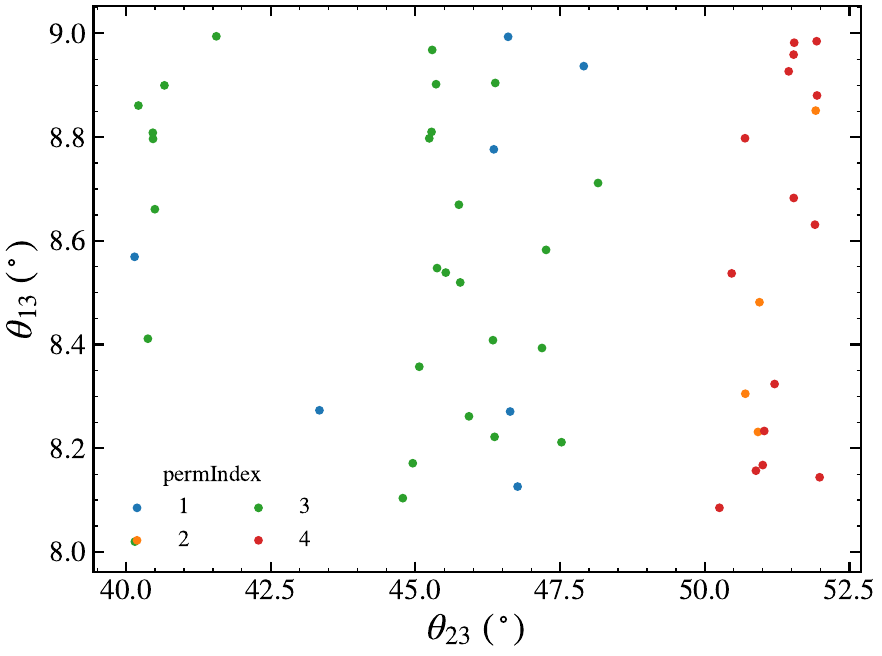}
\includegraphics[width=0.48\linewidth]{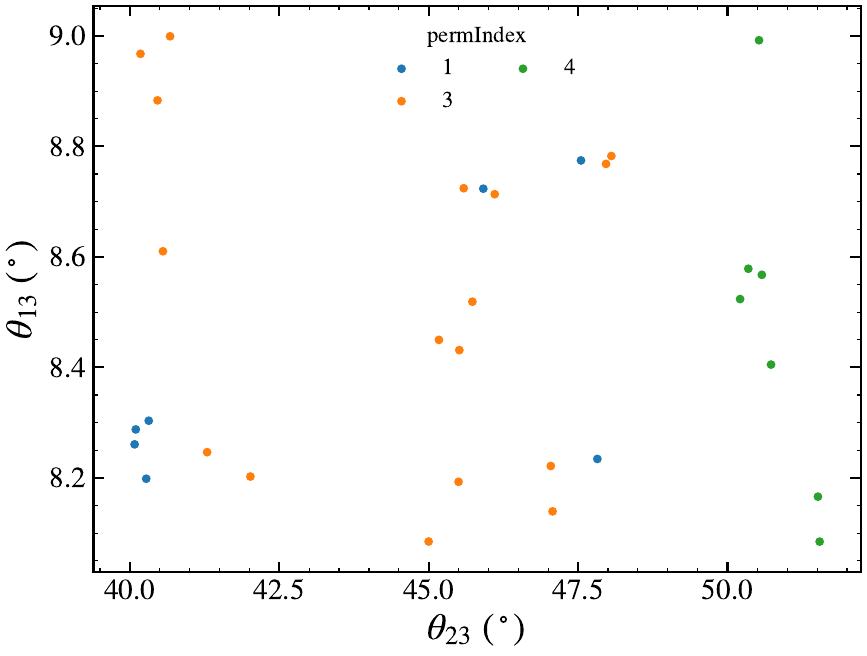}
\caption{$\theta_{23}$ Vs. $\theta_{13}$ for NO with $k_N=-1$ and $r_E=(3,-3,-5)$ at $\tau_1$ (left) and $\tau_2$ (right). Colors indicate the charged lepton permutation index.}
\label{fig:theta23_theta13}
\end{figure}

\begin{figure}[t]
\centering
\includegraphics[width=0.48\linewidth]{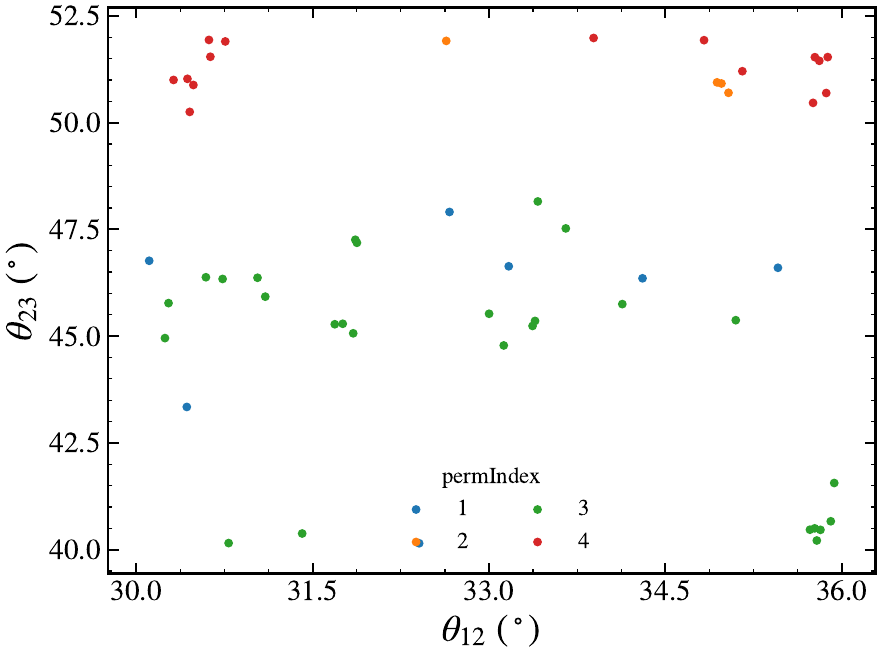}
\includegraphics[width=0.48\linewidth]{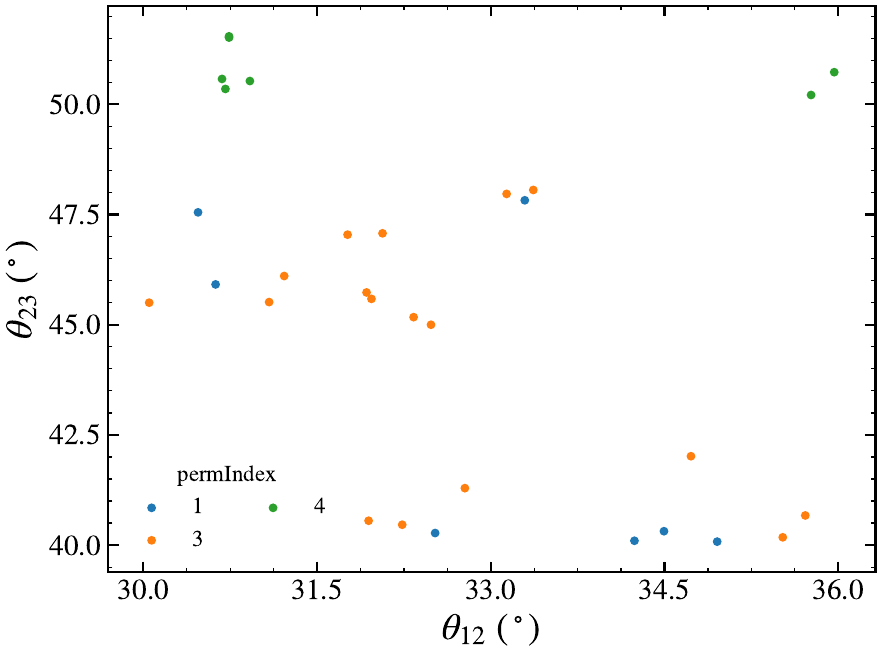}
\caption{$\theta_{12}$ Vs. $\theta_{23}$  for NO with $k_N=-1$ and $r_E=(3,-3,-5)$ at $\tau_1$ (left) and $\tau_2$ (right). Colors indicate the charged lepton permutation index.}
\label{fig:theta12_theta23}
\end{figure}

The allowed points show clear correlations in the mixing angle planes, especially in the $(\theta_{23},\theta_{13})$ projection, where the model predicts narrow bands rather than wide spread out regions. This indicates that once $\tau$ and the charged lepton modular weights are fixed, the neutrino phases cannot vary arbitrarily but are strongly constrained by the observed PMNS structure. The mixing angle correlations are shown in
Figs.~\ref{fig:theta23_theta13}-\ref{fig:theta12_theta23}.

\begin{figure}[t]
\centering
\includegraphics[width=0.48\linewidth]{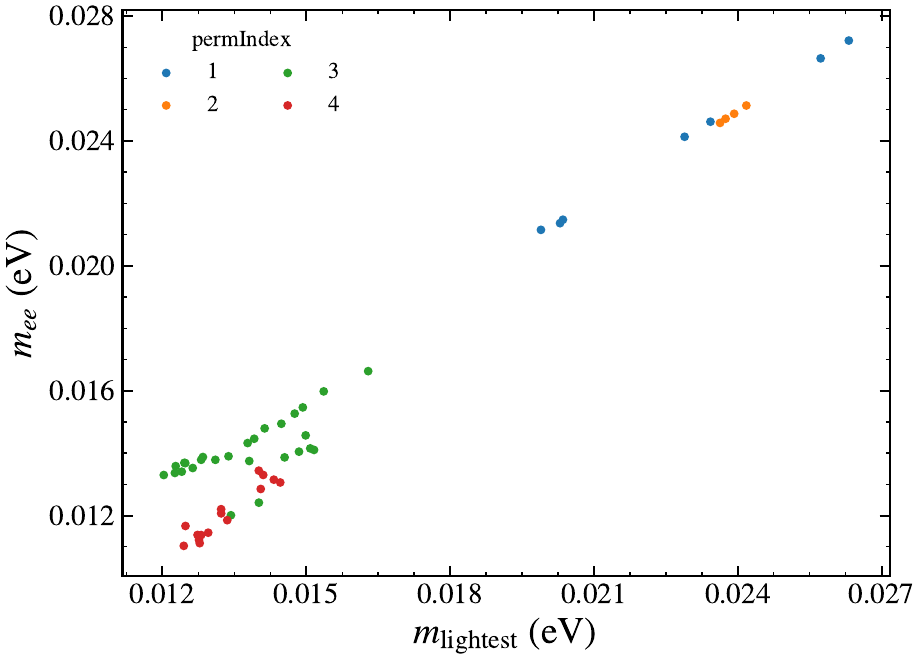}
\includegraphics[width=0.48\linewidth]{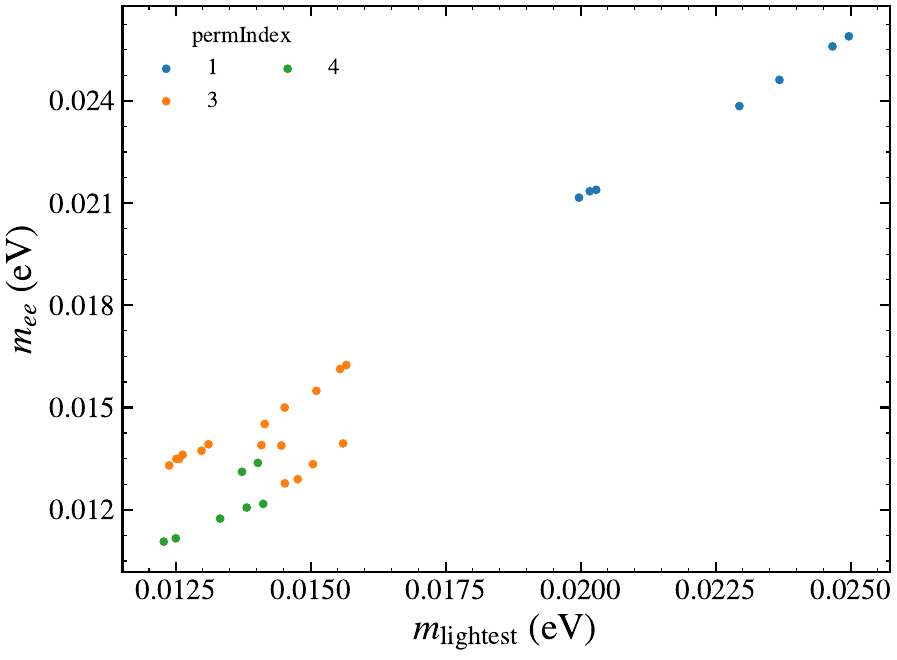}
\caption{Correlation between $m_{ee}$ and the lightest neutrino mass for $\tau_1$ (left) and $\tau_2$ (right), with $k_N=-1$ and $r_E=(3,-3,-5)$. Colors indicate permutation index.}
\label{fig:mee_mlight}
\end{figure}

\begin{figure}[t]
\centering
\includegraphics[width=0.48\linewidth]{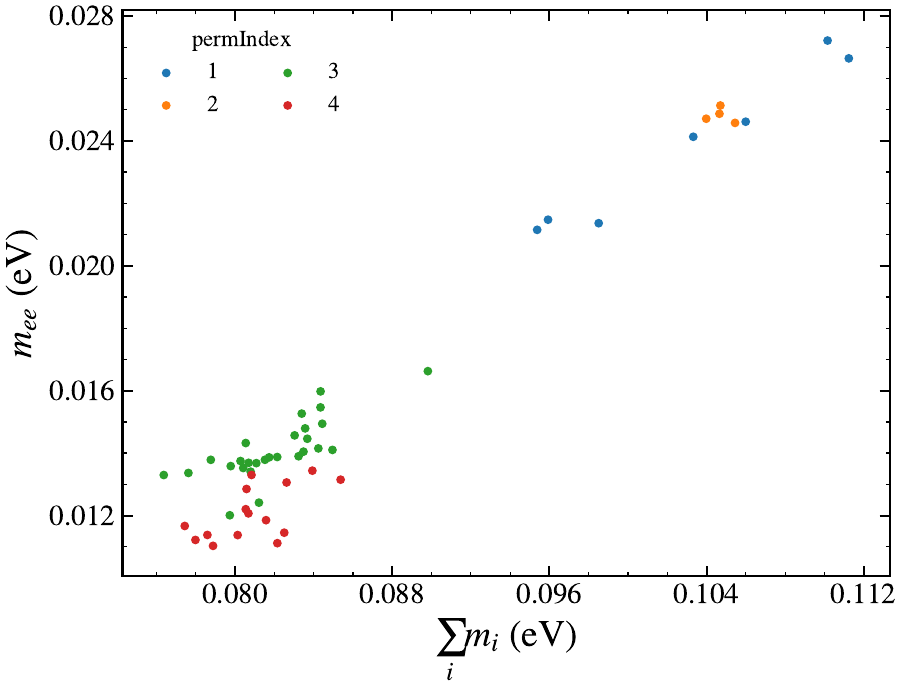}
\includegraphics[width=0.48\linewidth]{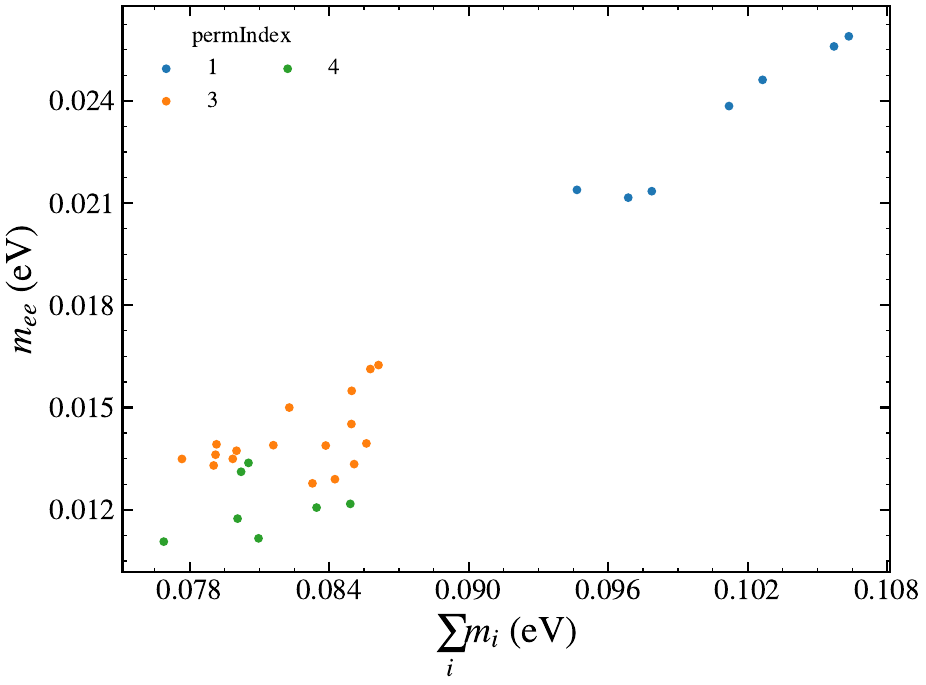}
\caption{Correlation between $m_{ee}$ and the sum of neutrino mass  for $\tau_1$ (left) and $\tau_2$ (right), with $k_N=-1$ and $r_E=(3,-3,-5)$. Colors indicate permutation index.}
\label{fig:mee_sum}
\end{figure}

\begin{figure}[t]
\centering
\includegraphics[width=0.48\linewidth]{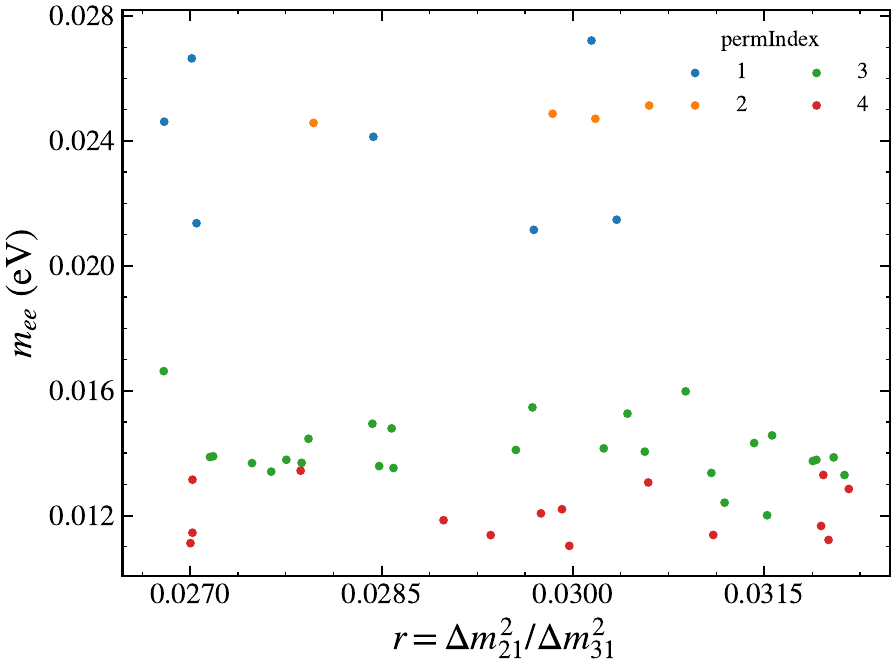}
\includegraphics[width=0.48\linewidth]{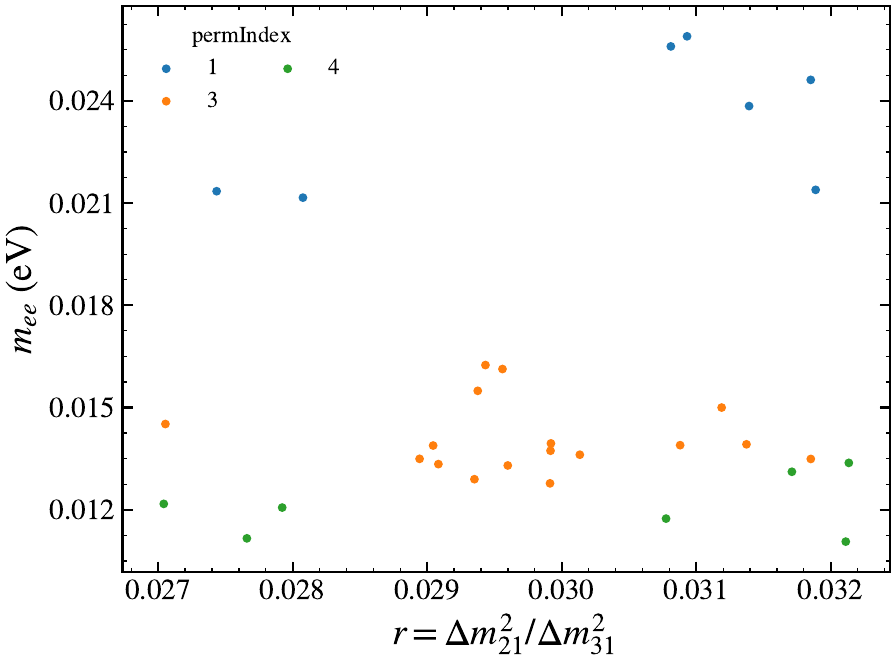}
\caption{Correlation between $m_{ee}$ and the neutrino mass square ratio $r=\frac{\Delta m_{12}^2}{\Delta m_{31}^2}$ for $\tau_1$ (left) and $\tau_2$ (right), with $k_N=-1$ and $r_E=(3,-3,-5)$. Colors indicate permutation index.}
\label{fig:mee_r}
\end{figure}

Such case also provides nontrivial predictions for the neutrino mass observables. The allowed points fill limited regions in the $(m_{ee},m_{\rm lightest})$, $(m_{ee},\Sigma)$ and $(m_{ee},r)$ planes, thereby the model correlates neutrinoless double beta decay and cosmological observables. In particular, the allowed parameter space does not fill the full normal ordering band, but occupies a smaller region selected by the modular structure and the phase constraints.
The predictions for neutrinoless double beta decay are shown in
Figs.~\ref{fig:mee_mlight}, \ref{fig:mee_sum}, \ref{fig:mee_r}.

We can conclude that although the three modular weight assignments of charged leptons can reproduce the same mass pattern for the same values of $\tau_{1,2}$, they produce different predictions of the neutrino phenomenological aspects. One may suggest that the charged lepton sector modulates neutrino parameters via the diagonalization matrix $U_e$, which controls the relative alignments between the charged lepton and neutrino sectors.

The solutions for the assignment $(3,-3,-5)$ are arisen with more than one charged lepton permutation, while in the assignments $(-3,-5,3)$ and $(-5,3,-3)$ they are more restrictive admitting only a single viable permutation. This pattern indicates that the model selects a preferred modular weight for the right handed neutrinos, as well as specific charged lepton alignments.
This is an impressive result from the point of view of modular flavour symmetry. As a result, the model shows predictivity in the continuous space of phase parameters, and also in the discrete structure of flavor alignments.



\section{Discussion}\label{sec:discussion}
The numerical analysis indicates highly constrained structure emerging from the non-holomorphic modular $A_4$ framework with universal couplings. Although the scan covers a broad range of phase parameters, the viable region remains rather limited, indicating that the combination of modular symmetry and universal couplings leads to strong predictivity. Several key features can be identified from the results:

\begin{itemize}

\item A notable outcome of the scan is the clear preference for the modular weight assignment $k_N=-1$. For other values of $k_N$, the model either fails to reproduce the observed mixing angles or cannot accommodate the ratio $r=\Delta m_{21}^2/\Delta m_{3\ell}^2$. This behavior can be understood from the modular weight dependence of the Dirac and Majorana mass matrices. In the case $k_N=-1$, the resulting structure of the seesaw mass matrix naturally leads to the correct hierarchy and mixing pattern. In this sense, the model effectively selects a preferred modular weight configuration, enhancing its predictive power. Furthermore, the assumption of universal couplings imposes additional constraints on the model, such that not all values of $k_N$ are compatible with the observed data.

\item The analysis also indicates a clear preference for normal neutrino mass ordering. Inverted ordering is either not realized or appears only in a very limited region of parameter space. This behavior can be understood from the limitations imposed by the universal coupling assumption. In this framework, since all couplings have equal magnitudes, the neutrino mass ordering is determined by constrained phases and one modular weight. These constraints restrict the model in such a way that not all ordering scenarios can be realized. This feature therefore appears to be an intrinsic property of the model rather than the result of parameter tuning.

\item A significant aspect is the presence of nontrivial correlations between the phases. Since the magnitudes of the couplings are fixed, the phases become the main source of flavor structure. The allowed solutions are therefore restricted to specific regions in parameter space. These correlations are responsible for  reproducing neutrino observables simultaneously, indicating the restrictive nature of the framework.

\item One of the most interesting outcomes of the analysis is the appearance of charged lepton permutation selection. Although the ordering of the charged lepton eigenvectors is, in principle, arbitrary, the data allow only a restricted set of permutations. This means that the model dynamically favors a particular relative alignment between $U_e$ and $U_\nu$. The fact that this pattern persists for different values of $\tau$ and for all charged lepton modular weight assignments considered here strongly suggests that it is a genuine structural feature of the framework rather than a numerical accident.


\item One of the important prediction of the model is the existence of a nonzero lower bound on the effective Majorana mass $m_{ee}$. The effective Majorana mass $m_{ee}$ involves a coherent sum of complex contributions proportional to $U_{ei}^2 m_i$. Although it is given by the modulus of this sum, the relative phases among the terms allow for possible cancellations. In the present framework, the universal couplings and restricted phase structure prevents complete destructive interference, leading to a nonvanishing lower bound on $m_{ee}$. These values can be tested in upcoming neutrinoless double beta decay experiments~\cite{nEXO:2017nam,LEGEND:2017cdu,KamLAND-Zen:2016pfg,CUPID:2019imh,SNO+:2015lza}.

\item Analysis also finds a strong correlation between $m_{ee}$ and the sum of neutrino masses $\sum_i m_i$. They arise directly from the neutrino mass spectrum produced by the modular structure. The obtained values of $\sum_i m_i$ are close to the cosmological limit, $\sum_i m_i \lesssim 0.12~\text{eV}$, obtained from Planck data \cite{Planck:2018vyg} combined with large scale structure observations. Thus, future measurements of the cosmic microwave background and the large scale structure will serve as a powerful and complementary probe of this model. 

\item These results show that the non-holomorphic modular $A_4$ framework with universal couplings is highly predictive. The model selects a specific modular weight, favors normal mass ordering, and predicts a nonzero value of $m_{ee}$, together with clear correlations among neutrino observables. These features distinguish it from more generic flavor models and make it testable in current and future experiments.
\end{itemize}

\section{Conclusion}\label{sec:conclusion}
In this work, we have investigated a non-holomorphic modular $A_4$ framework for lepton flavor under the assumption of universal coupling magnitudes. In this setting, the hierarchy of charged lepton masses arises directly from the modular structure, without introducing hierarchical Yukawa parameters or requiring fine tuning. By fixing the modulus $\tau$ through the charged lepton sector, we identified viable solutions clustered near modular fixed points. This suggests that the observed mass hierarchy can be interpreted as a consequence of small deviations from underlying residual symmetries.

We then extended the analysis to the neutrino sector through the type-I seesaw mechanism. The couplings are assumed to be equal in magnitude up to relative phases. We carried out a numerical study over the remaining phase parameters and the modular weight $k_N$ of the right handed neutrinos. A central outcome is the high level of predictivity of the model: once $\tau$ is fixed, the neutrino data select a unique value, $k_N = -1$, and viable solutions are obtained only for normal mass ordering. This greatly restricts the parameter space and leads to pronounced correlations among the physical observables.

The model further predicts nontrivial correlations among the mixing angles, the mass squared ratio $r$, the effective Majorana mass $m_{ee}$, and the total neutrino mass $\Sigma m_i$. The viable solutions are confined to relatively narrow regions of parameter space, and a nonzero lower bound on $m_{ee}$ emerges naturally. This places a portion of the parameter space within the reach of upcoming neutrinoless double beta decay experiments. 

Overall, we conclude that the non-holomorphic modular $A_4$ model simply describe lepton masses and mixing in the case of universal couplings. The model is testable and predictive, and it is free from ad hoc hierarchies or tuning of parameters, making it an attractive candidate for understanding the origin of lepton masses and mixing.



\section{Acknowledgment} 
This work was funded by the Deanship of Graduate Studies and Scientific Research at Jouf University
under grant No. (DGSSR-2025-02-01638)



\begin{thebibliography}{99}

 
 \bibitem{modulargroups}
  R.~de Adelhart Toorop, F.~Feruglio and C.~Hagedorn,
  Nucl.\ Phys.\ B {\bf 858}, 437 (2012)
  doi:10.1016/j.nuclphysb.2012.01.017
  [arXiv:1112.1340 [hep-ph]].

\bibitem{Feruglio:2017spp}
  F.~Feruglio,
  doi:$10.1142/9789813238053_0012$
  arXiv:1706.08749 [hep-ph].






\bibitem{Qu:2024rns}
B.~Y.~Qu and G.~J.~Ding,
JHEP \textbf{08}, 136 (2024)
doi:10.1007/JHEP08(2024)136
[arXiv:2406.02527 [hep-ph]].

\bibitem{Okada:2025jjo}
H.~Okada and Y.~Orikasa,
[arXiv:2501.15748 [hep-ph]].

\bibitem{Loualidi:2025tgw}
M.~A.~Loualidi, M.~Miskaoui and S.~Nasri,
Phys. Rev. D \textbf{112}, no.1, 1 (2025)
doi:10.1103/1py2-cmfx
[arXiv:2503.12594 [hep-ph]].

\bibitem{Abbas:2025nlv}
M.~Abbas,
PHEP \textbf{2025}, 7 (2025)
doi:10.31526/PHEP.2025.07

\bibitem{Nomura:2025raf}
T.~Nomura and H.~Okada,
Chin. Phys. \textbf{50}, no.2, 023108 (2026)
doi:10.1088/1674-1137/ae15ee
[arXiv:2506.02639 [hep-ph]].

\bibitem{Zhang:2025dsa}
X.~Zhang and Y.~Reyimuaji,
Phys. Rev. D \textbf{112}, no.7, 075050 (2025)
doi:10.1103/17p3-bw5r
[arXiv:2507.06945 [hep-ph]].

\bibitem{Priya:2025wdm}
Priya, L.~Singh, B.~C.~Chauhan and S.~Verma,
JHEP \textbf{01}, 036 (2026)
doi:10.1007/JHEP01(2026)036
[arXiv:2508.05047 [hep-ph]].

\bibitem{Kumar:2025nut}
B.~Kumar and M.~K.~Das,
Phys. Lett. B \textbf{872}, 140064 (2026)
doi:10.1016/j.physletb.2025.140064
[arXiv:2509.01205 [hep-ph]].

\bibitem{Nanda:2025lem}
S.~K.~Nanda, M.~Ricky Devi and S.~Patra,
[arXiv:2509.22108 [hep-ph]].

\bibitem{Jangid:2025thp}
S.~Jangid and H.~Okada,
[arXiv:2510.17292 [hep-ph]].

\bibitem{Gao:2025jlw}
X.~Y.~Gao and C.~C.~Li,
[arXiv:2512.07158 [hep-ph]].

\bibitem{Tapender:2026ets}
Tapender and S.~Verma,
[arXiv:2602.17243 [hep-ph]].

\bibitem{Majhi:2026jdk}
R.~Majhi, M.~K.~Behera and R.~Mohanta,
[arXiv:2602.23018 [hep-ph]].











\bibitem{Nomura:2024atp}
T.~Nomura and H.~Okada,
[arXiv:2408.01143 [hep-ph]].

\bibitem{Nomura:2024vzw}
T.~Nomura, H.~Okada and O.~Popov,
Phys. Lett. B \textbf{860}, 139171 (2025)
doi:10.1016/j.physletb.2024.139171
[arXiv:2409.12547 [hep-ph]].

\bibitem{Ding:2024inn}
G.~J.~Ding, J.~N.~Lu, S.~T.~Petcov and B.~Y.~Qu,
[arXiv:2408.15988 [hep-ph]].

\bibitem{Li:2024svh}
C.~C.~Li, J.~N.~Lu and G.~J.~Ding,
JHEP \textbf{12}, 189 (2024)
doi:10.1007/JHEP12(2024)189
[arXiv:2410.24103 [hep-ph]].


\bibitem{Zhang:2026kyy}
X.~Zhang and Y.~Reyimuaji,
[arXiv:2603.19104 [hep-ph]].

\bibitem{Li:2025kcr}
C.~C.~Li and G.~J.~Ding,
JHEP \textbf{01}, 032 (2026)
doi:10.1007/JHEP01(2026)032
[arXiv:2509.15183 [hep-ph]].

\bibitem{pdg}
 S. Navaset al.(Particle Data Group), Phys. Rev. D110, 030001 (2024) and 2025 update


\bibitem{Esteban:2018azc}
  I.~Esteban, M.~C.~Gonzalez-Garcia, A.~Hernandez-Cabezudo, M.~Maltoni and T.~Schwetz,
  JHEP {\bf 1901}, 106 (2019)
  doi:10.1007/JHEP01(2019)106
  [arXiv:1811.05487 [hep-ph]].

\bibitem{Goswami:2025jde}
S.~T.~Goswami and S.~Roy,
Nucl. Phys. B \textbf{1022}, 117275 (2026)
doi:10.1016/j.nuclphysb.2025.117275
[arXiv:2501.18181 [hep-ph]].

\bibitem{Hagedorn:2017zks}
C.~Hagedorn,
[arXiv:1705.00684 [hep-ph]].

\bibitem{Altarelli:2010gt}
G.~Altarelli and F.~Feruglio,
``Discrete Flavor Symmetries and Models of Neutrino Mixing,''
Rev.\ Mod.\ Phys.\ \textbf{82}, 2701 (2010),
arXiv:1002.0211 [hep-ph].

\bibitem{King:2013eh}
S.~F.~King and C.~Luhn,
``Neutrino Mass and Mixing with Discrete Symmetry,''
Rept.\ Prog.\ Phys.\ \textbf{76}, 056201 (2013),
arXiv:1301.1340 [hep-ph].

\bibitem{Feruglio:2012cw}
F.~Feruglio, C.~Hagedorn, and R.~Ziegler,
``Lepton Mixing Parameters from Discrete and CP Symmetries,''
JHEP \textbf{07}, 027 (2013),
arXiv:1211.5560 [hep-ph].

\bibitem{Holthausen:2012dk}
M.~Holthausen, M.~Lindner, and M.~A.~Schmidt,
``CP and Discrete Flavour Symmetries,''
JHEP \textbf{04}, 122 (2013),
arXiv:1211.6953 [hep-ph].

\bibitem{Kobayashi:2018scp}
T.~Kobayashi \textit{et al.},
``Modular $A_4$ invariance and neutrino mixing,''
JHEP \textbf{11}, 196 (2018),
arXiv:1808.03012 [hep-ph].

\bibitem{Criado:2018thu}
J.~C.~Criado and F.~Feruglio,
``Modular invariance faces precision neutrino data,''
SciPost Phys.\ \textbf{5}, 042 (2018),
arXiv:1807.01125 [hep-ph].

\bibitem{King:2020qaj}
S.~F.~King and Y.~L.~Zhou,
``Modular symmetry and predictions for lepton mixing,''
JHEP \textbf{04}, 2021,
arXiv:2005.06897 [hep-ph].

\bibitem{nEXO:2017nam}
J.~B.~Albert \textit{et al.} [nEXO Collaboration],
doi:10.1103/PhysRevC.97.065503,
arXiv:1710.05075 [nucl-ex].

\bibitem{LEGEND:2017cdu}
N.~Abgrall \textit{et al.} [LEGEND Collaboration],
doi:10.1063/1.5007652,
arXiv:1709.01980 [physics.ins-det].

\bibitem{KamLAND-Zen:2016pfg}
A.~Gando \textit{et al.} [KamLAND-Zen Collaboration],
doi:10.1103/PhysRevLett.117.082503,
arXiv:1605.02889 [hep-ex].

\bibitem{CUPID:2019imh}
G.~Wang \textit{et al.} [CUPID Interest Group],
arXiv:1907.09376 [physics.ins-det].

\bibitem{SNO+:2015lza}
S.~Andringa \textit{et al.} [SNO+ Collaboration],
doi:10.1155/2016/6194250,
arXiv:1508.05759 [physics.ins-det].

\bibitem{Planck:2018vyg}
N.~Aghanim \textit{et al.} [Planck Collaboration],
``Planck 2018 results. VI. Cosmological parameters,''
Astron.\ Astrophys.\ \textbf{641}, A6 (2020),
arXiv:1807.06209 [astro-ph.CO].

\end{thebibliography}
\end{document}